\documentclass[%
  reprint,
  superscriptaddress,
  amsmath,amssymb,
  aps,
  prl,
]{revtex4-2}

\usepackage{graphicx}
\usepackage{bm}
\usepackage{mathtools}
\usepackage{xcolor}
\usepackage{cancel}
\usepackage{dcolumn}

\usepackage[colorlinks=true,allcolors=blue]{hyperref}

\def\be{\begin{equation}}
\def\ee{\end{equation}}
\def\bea{\begin{eqnarray}}
\def\eea{\end{eqnarray}}
\def \dd {\mathrm{d}}

\begin{document}

\title{Kerr Black Hole Ringdown in Effective Field Theory}

\author{William L. Boyce}
\email{wlb26@cam.ac.uk}
\affiliation{Department of Applied Mathematics and Theoretical Physics, University of Cambridge, Wilberforce Road, Cambridge CB3 0WA, United Kingdom}

\author{Jorge E. Santos}
\email{jss55@cam.ac.uk}
\affiliation{Department of Applied Mathematics and Theoretical Physics, University of Cambridge, Wilberforce Road, Cambridge CB3 0WA, United Kingdom}

\date{March 12, 2026}

\begin{abstract}
We develop a systematic effective field theory calculation of the quasinormal modes of Kerr black holes valid for arbitrary spin, providing model-independent corrections to their ringdown spectrum directly relevant for gravitational-wave observations. Close to extremality, the effective field theory corrections in the grand‑canonical ensemble exhibit an oscillatory dependence on $\log \tau_{H}$, with $\tau_H \equiv T_H/\Omega_H$ a dimensionless measure of the black hole temperature. This behaviour signals an underlying discrete‑scale‑invariant structure.
\end{abstract}

\maketitle

{\it Introduction.} General Relativity (GR) continues to exhibit remarkable success in accounting for the rapidly expanding body of observational data produced by the LIGO–Virgo–KAGRA (LVK) Collaboration during its recent observing runs \cite{GWTC4_2025}. At the same time, black‑hole physics has consistently driven progress in our understanding of gravity, and the increasing precision of compact‑object observations promises to sharpen this role even further. Yet even with these successes, GR cannot be the final word: it predicts physical singularities \cite{Penrose1965}, offers no mechanism to reconcile gravity with quantum field theory \cite{tHooftVeltman1974}, and leaves unresolved both the origin of black‑hole entropy \cite{Bekenstein1973,Wald2025} and the fate of information in black‑hole evaporation \cite{Hawking1975}. These long‑standing issues point inexorably toward the need for a quantum theory of gravity.

In view of these open problems, we adopt an agnostic stance toward the ultraviolet completion of gravity and work within the framework of effective field theory (EFT). EFT provides the most general and systematically improvable description of low‑energy deviations from GR, encoding all allowed corrections as higher‑dimensional operators suppressed by a heavy scale and constrained by locality, unitarity, Lorentz invariance, analyticity, and, crucially for gravity, diffeomorphism invariance. These principles ensure that any extension of Einstein gravity remains predictive \cite{Figueras:2024bba,Gavassino:2026xjw} while keeping assumptions about the underlying UV theory to a minimum. At the same time, EFT has well‑defined limits: it cannot capture intrinsically non‑local proposals—such as island constructions \cite{Almheiri:2019qdq,Almheiri:2019psf,Almheiri:2019yqk,Penington:2019kki,Almheiri:2019hni,Almheiri:2019psy}, fuzzball microstate structure \cite{Mathur:2005zp,Bena:2022fuzzballs,Bena:2022snowmass,Mayerson:2020fuzzballs,Bena:2025microstate,Heidmann:2020holomorphic,Shigemori:2019superstrata,Heidmann:2025pbb}, or other horizon‑scale non‑local effects—which lie outside the domain of validity of a local derivative expansion.

Although EFT coefficients are often assumed to be minuscule—suppressed by appropriate powers of the Planck scale—their actual size depends sensitively on the spectrum of heavy fields that has been integrated out. As a result, these coefficients encode valuable information about possible ultraviolet matter sectors and can place meaningful constraints on hypothetical new physics. Moreover, their impact is not uniformly small: near extremality, parity‑even higher‑derivative operators can generate parametrically enhanced corrections, producing sizeable deviations from General‑Relativistic predictions while remaining fully consistent with EFT power counting \cite{Horowitz:2023xyl,Horowitz:2024dch}.

In this work, we examine an important component of this broader effort: the quasinormal‑mode (QNM) spectrum of rotating black holes, of the type routinely observed by the LVK Collaboration. The EFT description of Schwarzschild black holes—including their QNMs—was pioneered in \cite{Cardoso:2018ptl}, whereas the EFT description of the QNM spectra of Reissner–Nordström black holes was developed only recently in \cite{Boyce:2025fpr}. However, much of the existing literature incorporates beyond‑GR corrections using a small‑spin expansion \cite{Cano:2020ringing,Cano:2021gravitational,Cano:2023univTeuk,Cano:2023higherderivativeQNM,Cano:2024ezp,Aly:2026otj} \footnote{A notable exception is \cite{Miguel:2023rzp}, which studies the QNMs of electromagnetic waves in an EFT framework on a fixed Kerr background.}. We show that this expansion breaks down once the dimensionless spin exceeds a critical value. For the mode under consideration, the breakdown occurs close to the point where the specific heat of a Kerr black hole changes sign, which is also precisely where the asymptotic expansion already shows clear signs of failure. This behaviour is likely due to the inherently asymptotic nature of small‑spin expansions. Although such failures are always possible for asymptotic series, it is the direct numerical comparison with the small‑spin expansion that makes the breakdown unmistakable \footnote{Our numerical data lie well outside the error bars reported by \cite{Cano:2024ezp}.}. These results highlight that accurate modelling of high‑spin black‑hole spectroscopy—precisely the regime probed by many LVK detections—requires techniques that extend beyond the traditional low‑spin approximation.

We further note that approaches relying solely on the eikonal limit introduce additional subtleties \cite{Cano:2024modifiedTeukolsky,Cano:2024HDGQNM,Cano:2025isospecEFT,Cano:2025eikonalEFT}. While the eikonal approximation captures certain high‑angular‑momentum features, subleading corrections—including those induced by higher‑derivative interactions—can contribute at the same order as the leading eikonal term, undermining the reliability of the expansion (see Sec.~4.8 of \cite{Reall:2021voz} for a detailed discussion). Consequently, the eikonal regime can become overly sensitive to the structure of the underlying EFT, and careless extrapolation may lead to qualitatively incorrect conclusions about the QNM spectrum. Understanding these limitations is therefore essential when using QNMs to test or constrain extensions of GR.

Finally, we emphasise that the methods developed in this work rely solely on the EFT framework and require no assumptions about the ultraviolet completion of gravity. Our analysis remains valid across the entire range of black‑hole spins, including the high‑spin regime where traditional approximations fail. This ensures that our results provide a fully controlled and model‑independent characterisation of quasinormal‑mode physics within the domain of applicability of EFT.

{\it The EFTs.} The most general EFT extension of the Einstein-Hilbert action involving up to eight derivatives is \cite{Endlich:2017tqa}
\begin{widetext}
\begin{equation}
\label{eq:GREFTB}
S = \frac{1}{2\kappa_{\rm 4D}^2}\int \sqrt{-g}\,\Big[
R 
+ d_1\,\kappa_{\rm 4D}^2\, R_{ab}{}^{cd} R_{cd}{}^{ef} R_{ef}{}^{ab}
+ d_2\,\kappa_{\rm 4D}^2\, R_{ab}{}^{cd} R_{cd}{}^{ef} \tilde{R}_{ef}{}^{ab}
+ d_3\,\kappa_{\rm 4D}^4\,\mathcal{C}^2
+ d_4\,\kappa_{\rm 4D}^4\,\tilde{\mathcal{C}}^2
+ d_5\,\kappa_{\rm 4D}^4\,\mathcal{C}\,\tilde{\mathcal{C}}
\Big] \, .
\end{equation}
\end{widetext}
Here $\kappa_{\rm 4D}^2 = 8\pi G_{\rm 4D}$, with $G_{\rm 4D}$ Newton’s constant. The dual Riemann tensor is defined as 
$\tilde{R}_{abcd} = \varepsilon_{abef} R^{ef}{}_{cd}$, where $\varepsilon_{abcd}$ is the volume form. We also define the Kretschmann scalar $\mathcal{C} = R_{abcd} R^{abcd}$ and its dual $\tilde{\mathcal{C}} = R_{abcd} \tilde{R}^{abcd}$. The equations of motion derived from this action are presented in the Supplemental Material (cf.\ Eq.~\ref{eq: EFT Einstein eqn}). The coefficients $d_K$ are Wilsonian parameters of energy dimension $-2$; that is, in a tree-level completion, $d_k \sim 1/\Lambda^2$, where $\Lambda$ is the energy scale of new physics. The terms proportional to $d_2$ and $d_5$ are parity odd, whereas the remaining terms preserve parity.

{\it Methodology.} We seek the EFT corrections to the QNM spectrum of Kerr black holes. This requires, first, determining the EFT corrections to the Kerr background itself. Because the higher-derivative terms in Eq.~(\ref{eq:GREFTB}) render Kerr no longer an exact solution, we treat the coefficients $d_i$ as infinitesimal, so that the EFT-corrected geometry is perturbatively close to Kerr. Perturbation theory then suffices to compute the leading corrections. This strategy has been successfully applied in related contexts \cite{Reall:2019sah,Horowitz:2024dch,Fernandes:2025vxg}, where EFT-corrected Kerr and Kerr–Newman solutions were obtained. In this first step, we search for stationary and axisymmetric black hole solutions, characterised by a stationary Killing vector $\partial_t$ and an axisymmetric Killing vector $\partial_\phi$ with $\phi \sim \phi + 2\pi$. A convenient parametrisation is
\begin{equation} \label{eq:EFTKmetricmain} \begin{aligned} \dd s^2_{\text{EFTK}} &= - \frac{\Delta_{\rm K}(r)}{\Sigma(r,x)}\, F_1(r,x)\,\big[\dd t - (1-x^2) F_4(r,x)\,\dd\phi\big]^2 \\ &\quad + \frac{1-x^2}{\Sigma(r,x)}\, F_3(r,x)\,\big[F_4(r,x)\, \dd t - (r^2 + a_{\rm K}^2)\, \dd\phi\big]^2 \\ &\quad + \Sigma(r,x)\, F_2(r,x)\left[\frac{\dd r^2}{\Delta_{\rm K}(r)} + \frac{\dd x^2}{1-x^2}\right]\, .
\end{aligned}
\end{equation}
Here $\Delta_{\rm K}(r) = r^2 - 2 M_{\rm K} r + a_{\rm K}^2$ and $\Sigma(r,x) = r^2 + a_{\rm K}^2 x^2$, with $|a_{\rm K}| \le M_{\rm K}$. The metric reduces to Kerr when $F_{1,2,3}(r,x) = F^{\rm K}_{1,2,3} = 1$ and $F_4(r,x) = F^{\rm K}_4 = a_{\rm K}$, in which case the spacetime coincides with a Kerr black hole of energy $E_K = 8\pi M_{\rm K}/\kappa_{\rm 4D}^2$ and angular momentum $J_K = a_{\rm K}\,E_{K}$. Accordingly, we expand
\begin{equation}
\label{eq:F expansion}
F_I(r,x) = F_I^{\rm K} + \sum_{k=1}^2 \frac{\kappa_{\rm 4D}^2}{M_{\rm K}^4}\,d_k\, f_I^{(k)}(r,x)+\sum_{k=3}^5 \frac{\kappa_{\rm 4D}^4}{M_{\rm K}^6}\, d_k\,f_I^{(k)}(r,x)\,,
\end{equation}
for $I = 1,\ldots,4$, and solve perturbatively for the functions $f_I^{(k)}$ (the integration procedure follows \cite{Horowitz:2024dch}). In the numerical procedure described in \cite{Horowitz:2024dch}, the black hole horizon, where $\Delta_K(r_+)=0$ vanishes, is held fixed under the EFT deformation, as is the periodicity of $\phi$. By imposing suitable boundary conditions at asymptotic infinity, we likewise keep the ratio $J/E^2=J_{K}/E^2_{K}$ fixed. All remaining thermodynamic quantities receive EFT corrections, and we reproduce the results of \cite{Reall:2019sah}. The odd-parity terms, proportional to $d_2$ and $d_5$, produce \emph{no} change in any thermodynamic quantity by symmetry, although they do generate non-trivial functions $f_I(r,x)$. Once these corrections are obtained, we proceed to the second step and compute the corresponding EFT‑deformed QNM spectrum. Because the EFT‑corrected black holes are stationary and axisymmetric, all perturbations can be decomposed into Fourier modes of the form $e^{-{\rm i}\omega t + {\rm i} m \phi}$ with $m \in \mathbb{Z}$. Imposing outgoing boundary conditions at asymptotic infinity and ingoing boundary conditions at the horizon—the defining boundary conditions for QNMs—turns the problem into a non‑Hermitian quadratic Sturm–Liouville eigenvalue problem, whose allowed frequencies $\omega$ are discrete. For Kerr black holes, perturbations \emph{separate} into radial and polar sectors, and the resulting spectrum can be organised by counting the nodes of the corresponding eigenfunctions. Along the polar direction, the number of zeros is $\ell - |m|$, implying $\ell \ge |m|$. Along the radial direction, the number of nodes defines the overtone index $n = 0, 1, \ldots$. The QNM spectrum may therefore be labelled as $\omega_{\ell m n}$. Under the map $\phi \mapsto - \phi$, the Kerr metric of angular momentum $J_K$ is pulled back to a metric in the Kerr family with angular momentum $-J_K$. Under this transformation, $m \mapsto - m$, so we use this to restrict to $J_K \geq 0$ while keeping $m \in \mathbb Z$. For all Kerr positive spins, the slowest‑decaying mode lies in the sector $\ell = m = 2$ \cite{Chandrasekhar1983,Leaver1985}. This is the sector we focus on when computing EFT corrections, which we define as
\begin{equation} \omega_{\ell m n} = \omega^{\rm K}_{\ell m n} + \sum_{k=1}^{2} \frac{\kappa_{\rm 4D}^{2}}{M_{\rm K}^{4}}\, d_k \, \delta\omega_{\ell m n}^{(k)} + \sum_{k=3}^{5} \frac{\kappa_{\rm 4D}^{4}}{M_{\rm K}^{6}}\, d_k \, \delta\omega_{\ell m n}^{(k)}\,,
\end{equation}
with $\{\omega^{\rm K}_{\ell m n}\}$ denoting the Kerr QNM spectrum. We determine the frequency shifts $\delta\omega_{\ell m n}^{\,(k)}$ through a sequence of steps, formalised in the Supplemental Material, following a variant of the method introduced in \cite{Cano:2021gravitational}—there applied only perturbatively in the spin, with the EFT limit obtained by considering finite $d_k$ and subsequently extrapolating to $d_k \to 0$. In our approach, the $d_k$ are taken perturbatively small from the outset (in the true spirit of EFT), while the spin is allowed to take any value consistent with the Kerr extremality bound, $|J_{\rm K}|/E_{\rm K}^{2} \le \kappa_{\rm 4D}^{2}/(8\pi)$. The main idea is as follows: One may write down a universal Teukolsky equation for any spacetime, even in the presence of arbitrary matter sources. In general, however, these equations cannot be solved in any practical way for a generic background or for generic forms of matter. The situation simplifies once the equations are linearised around an algebraically special spacetime: in that case one obtains two partial differential equations for two Newman–Penrose components of the Weyl tensor, each sourced by perturbations of the corresponding stress-energy tensor. For typical matter fields, the source terms themselves depend on metric perturbations in a complicated way, so the resulting system is not closed and therefore not solvable by standard decoupling methods. The notable exceptions are cases where the matter sector is sufficiently simple for a closed system to exist (for instance, in Kerr–Newman–(A)dS backgrounds). Our case presents two complications. First, the EFT‑corrected black holes are not algebraically special. Second, the source terms in the Teukolsky equations arise from the effective stress-energy tensor generated by the higher‑derivative corrections. The lack of algebraic speciality is not fatal: our backgrounds differ from Kerr only perturbatively in the small parameters $d_k$, so the departure from type~D is itself of order $d_k$ (possibly multiplied by Kerr perturbations). This deviation can therefore be treated as an additional \emph{source} for the Teukolsky equation. The second complication can be handled in much the same way. The effective stress-energy tensor entering the universal Teukolsky equation contributes only perturbatively and depends solely on perturbations of the background geometry—that is, on perturbations of Kerr itself. These effective sources can be constructed using the procedures of \cite{Wald:1978vm,Cohen:1974cm,COHEN19755,Chrzanowski:1975wv}, which express metric perturbations of Kerr (and more generally of any type~D spacetime) in terms of a pair of complex scalar Hertz potentials, $\Psi^{H}_{\pm}$, each of which satisfies a sourceless Teukolsky‑type equation. Furthermore, \cite{Wald:1978vm} showed that one needs only $\Psi^{H}_{-}$ or $\Psi^{H}_+$ to generate the most general metric perturbation of a Kerr black hole (up to trivial shifts in the mass and angular momentum). This is consistent with the expected counting of degrees of freedom: the Hertz potentials are complex scalars, whereas a four‑dimensional graviton carries two real propagating degrees of freedom. The familiar isospectrality of Kerr black holes can then be understood as the statement that, for Kerr, the real and imaginary parts of $\Psi^{H}_{-}$ (or equivalently of $\Psi^{H}_{+}$) obey the same QNM spectrum. In practice, however, working directly with the real and imaginary parts of the Hertz potentials leads to cumbersome expressions. It is therefore more convenient to work with perturbations generated \emph{simultaneously} by $\Psi^{H}_-$ and $\Psi^{H}_+$—which naively appears to double the number of degrees of freedom—while imposing the appropriate relations between their real and imaginary parts so that the physical content remains unchanged. We are therefore left with a system of two unknown fields, $\psi_0$ and $\psi_4$, obeying Teukolsky equations sourced by an effective (albeit intricate) stress-energy tensor constructed from the Hertz potentials $\Psi^{H}_{\pm}$. This coupled system can be solved systematically within \emph{degenerate} perturbation theory. After performing the perturbative expansion, the problem reduces to determining the QNM frequency shifts $\delta\omega_{\ell m n}$ and the mixing between the $\Psi^{H}_-$ and $\Psi^{H}_+$ sectors implied by the degeneracy structure. Because the degeneracy is two‑fold, each $(\ell,m,n)$ mode yields two distinct corrections, which we denote by $\delta\omega^{(k)\;\pm}_{\ell m n}$. The numerical procedure used to compute $\delta\omega^{(k)\;\pm}_{\ell m n}$ is described in detail in the Supplemental Material.

{\it Results.} As noted above, we focus on the EFT corrections to the least‑damped Kerr quasinormal mode, namely the $\ell=m=2$, $n=0$ mode, for all spins satisfying \footnote{Our numerical methods can also handle the case $j=j_{\rm ext}$, but we leave this analysis for future work.} $j \equiv |J_{\rm K}|/E_{\rm K}^{2} < \kappa_{\rm 4D}^{2}/(8\pi) \equiv j_{\rm ext}$. We also note that the Kerr black hole admits a $\mathbb{Z}_2$ symmetry, $x\to -x$, under which the odd‑parity corrections satisfy a characteristic relation (proved in the Supplemental Material), namely $\delta\omega^{(k)\;-}_{\ell m n} = -\delta\omega^{(k)\;+}_{\ell m n}$ (with $k=2,5$). Consequently, for odd‑parity deformations we display only $\delta\omega^{(k)\;+}_{\ell m n}$ in our figures. We compute the EFT‑deformed QNM frequencies at fixed energy and angular momentum; that is, we compare the QNM spectra of the Kerr black hole and its EFT‑deformed counterpart while holding the conserved charges fixed in the microcanonical ensemble. As emphasised in \cite{Reall:2019sah}, this ensemble is not well suited for analysing the approach to extremality, owing to the non‑analytic behaviour of certain thermodynamic quantities of Kerr. It is, nevertheless, the physically relevant choice for astrophysical applications of our results.

\begin{figure*}[!t]
    \includegraphics[width=\linewidth]{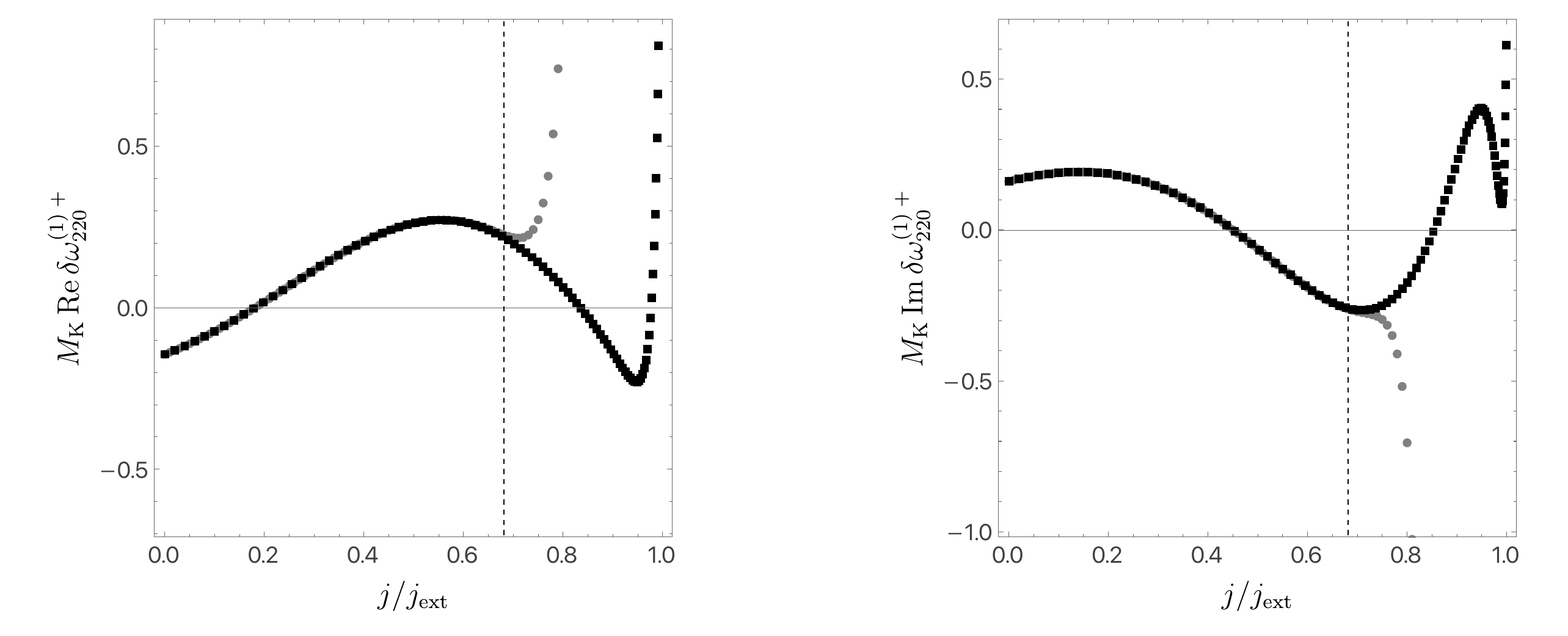}
    \caption{\label{fig:data1}$M_{\rm K}\,{\rm Re}\,\delta\omega^{(1)\;+}_{220}$ (left) and $M_{\rm K}\,{\rm Im}\,\delta\omega^{(1)\;+}_{220}$ (right) as functions of $j/j_{\rm ext}$. Black squares show our numerical results; grey discs reproduce the data of \cite{Cano:2024ezp}, publicly available in the \href{https://github.com/pacmn91/BeyondKerrQNM/tree/main?tab=readme-ov-file}{GitHub repository}. The vertical dashed line at $j/j_{\rm ext}=\sqrt{2\sqrt{3}-3}$ indicates the critical spin where the small‑spin expansion breaks down.}
\end{figure*}

Figure~\ref{fig:data1} shows $M_{\rm K}\,{\rm Re}\,\delta\omega^{(1)\;+}_{220}$ (left) and $M_{\rm K}\,{\rm Im}\,\delta\omega^{(1)\;+}_{220}$ (right) over the range $0 < j/j_{\rm ext} \leq 0.998686$. Note that $j/j_{\rm ext}=0.998686$ lies above the Thorne bound \cite{Thorne1974}. Corresponding plots for the remaining EFT deformations are provided in the Supplemental Material. The black squares denote our numerical results, while the gray discs reproduce the data of \cite{Cano:2024ezp}, publicly available in the \href{https://github.com/pacmn91/BeyondKerrQNM/tree/main?tab=readme-ov-file}{GitHub repository}. The black dashed line at $j/j_{\rm ext}=\sqrt{2\sqrt{3}-3}$ marks the critical spin where the small‑spin expansion breaks down. We have generated analogous plots for all other EFT deformations in Eq.~(\ref{eq:GREFTB}), and in every case we observe the same breakdown of the small‑spin expansion for spins $j/j_{\rm ext}>\sqrt{2\sqrt{3}-3}$. All the data generated in this work is available in the \href{https://github.com/jorgealberich/Effective-Field-Theory-of-Kerr-Black-Hole-Ringdown}{GitHub repository}.

We have also examined the behaviour of the spectrum in the grand‑canonical ensemble, where we compare configurations at fixed Hawking temperature $T_H$ and angular velocity $\Omega_H$. The QNM shifts are defined by
\begin{multline}
\varpi_{\ell m n} = \varpi^{\rm K}_{\ell m n} + \sum_{k=1}^{2} \kappa_{\rm 4D}^{2}\,\Omega_H^4\, d_k \, \Delta\varpi_{\ell m n}^{(k)}
\\
+ \sum_{k=3}^{5} \kappa_{\rm 4D}^{4}\Omega_H^6\, d_k \, \Delta\varpi_{\ell m n}^{(k)}\,,
\end{multline}
where $\varpi_{\ell m n}\equiv \omega_{\ell m n}/T_H$. For Kerr, the mode $\varpi_{220}$ is known to approach a constant in the extremal limit, implying that $\omega_{220}$ tends to zero as extremality is approached. Our goal is to determine how this behaviour is modified once higher‑derivative corrections are included in the EFT description. The results of \cite{Horowitz:2023xyl} strongly suggest that $\Delta\varpi_{220}^{(k)}$ remains finite in the extremal limit. This follows from the fact that the near‑horizon geometries constructed in \cite{Horowitz:2023xyl} possess an $SO(2,1)$ symmetry that enforces this behaviour and are completely smooth, even though the full spacetime is only mildly singular.

\begin{figure}[!t]
    \includegraphics[width=\linewidth]{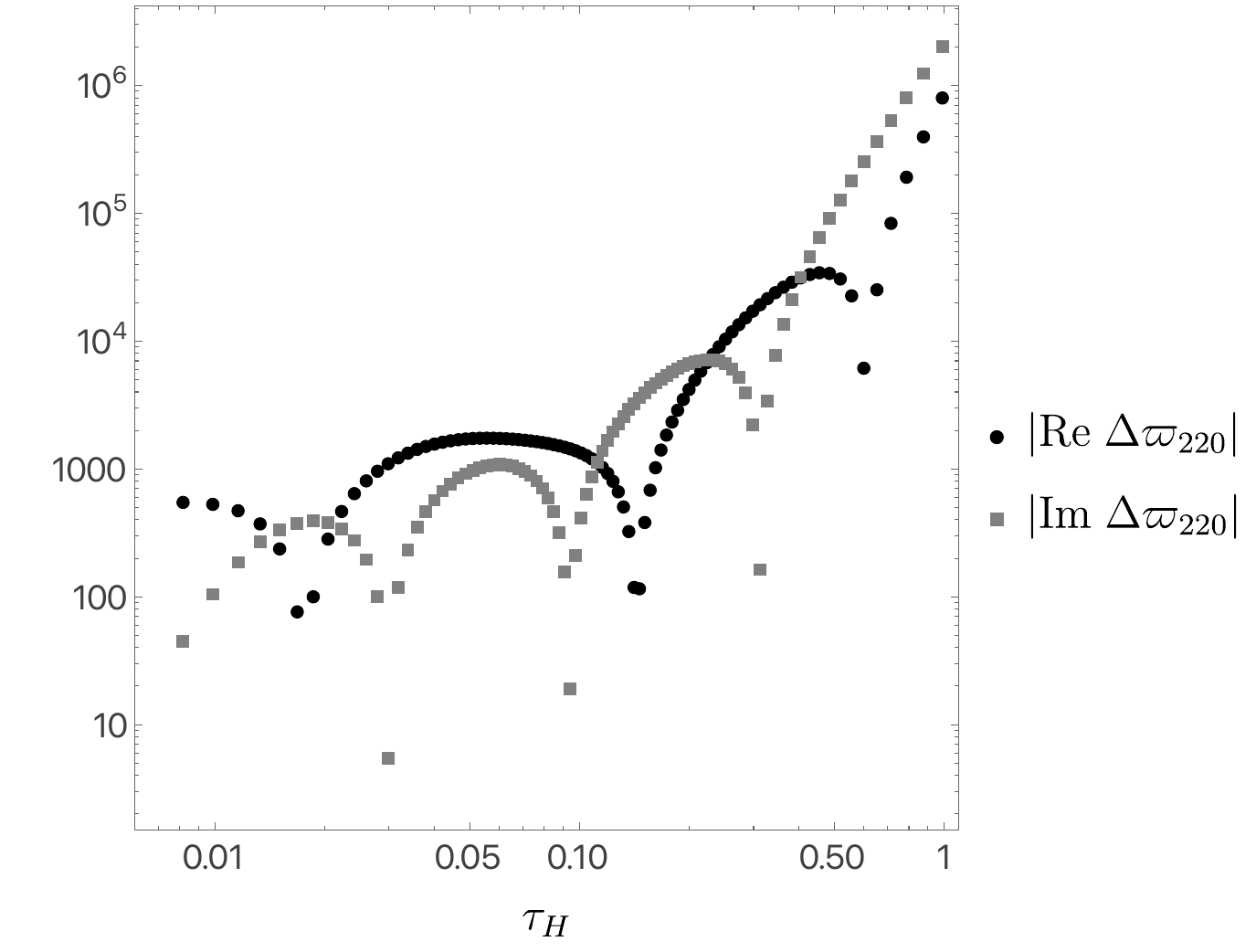}
    \caption{\label{fig:gran}Real (black disks) and imaginary (grey squares) parts of the QNM EFT shifts in the grand‑canonical ensemble versus $\tau_H \equiv T_H/\Omega_H$ (log–log scale). The curves develop logarithmic oscillations toward $\tau_H\to0$, with the real‑part ``echoing period'' roughly twice that of the imaginary part.}
\end{figure}

In Fig.~\ref{fig:gran} we show the real (black disks) and imaginary (grey squares) parts of the QNM EFT shifts in the grand‑canonical ensemble, for the same EFT deformations as in Fig.~\ref{fig:data1}, plotted as a function of $\tau_H \equiv T_H/\Omega_H$ on a log–log scale. Although normalising the EFT shifts by $\Omega_H$ is natural in the grand‑canonical ensemble, it introduces an artificial singular behaviour in the $\Omega_H\to0$ (or $\tau_H\to+\infty$) limit, as seen in Fig.~\ref{fig:gran}. As we approach $\tau_H=0$, the shifts exhibit an oscillatory pattern with an approximately fixed frequency in $\log \tau_H$. The remaining deformations display the same qualitative behaviour, although for the $k=3,4$ sectors the oscillations only become visible at lower values of $\tau_H$. The ``echoing period'' of the real part is approximately twice that of the imaginary part across all EFT sectors we have investigated.

{\it Discussion.} We have computed the EFT corrections to the QNM spectrum of rotating Kerr black holes for the first five non-trivial higher-derivative operators, working perturbatively in the Wilson coefficients $d_k$ but non-perturbatively in the dimensionless spin $j/j_{\rm ext}$. In all cases, we find that the small-spin expansion breaks down well before the extremal regime, suggesting that perturbation theory around slowly rotating solutions has a parametrically limited domain of validity. This breakdown reflects the limitations of the auxiliary expansion in spin rather than any failure of the EFT itself. By contrast, our computation remains uniformly controlled for all $0 \le j < j_{\rm ext}$ provided the Wilson coefficients are sufficiently small. The resulting data set provides the first controlled access to the QNM spectrum of EFT-deformed Kerr black holes at high spin, and may prove valuable in searches for new degrees of freedom for which EFT methods remain applicable—such as the presence of additional, possibly light, fields coupled to gravity.

A potential concern is the regime close to extremality, where certain thermodynamic quantities become non-analytic and higher-derivative corrections can be parametrically enhanced. In our analysis, all frequency shifts are computed consistently to linear order in $d_k$. For any fixed spin strictly below $j_{\rm ext}$, the perturbative expansion remains valid within the standard regime of applicability of EFT. While a complete understanding of the extremal limit would ultimately require control beyond linear order in $d_k$, our results provide the leading EFT corrections arbitrarily close to extremality in a controlled expansion.

It has recently been suggested that UV completeness might place bounds on the EFT corrections to black hole QNMs \cite{Melville:2024zjq}. In particular, there should be no \emph{measurable} increase in the decay lifetime $\theta = 1/| \text{Im}(\omega)|$ of these perturbations, which is to say that
\begin{equation}
|\text{Re}(\omega^{\rm K})|\,\delta \theta = \frac{|\text{Re}(\omega^{\rm K})|}{\text{Im}(\omega^{\rm K})^2}\,\text{Im}(\delta \omega) \lesssim 1 .
\end{equation}
We observe that $\text{Im}(\delta \omega_{220}^{(1)+})$ changes sign as $j$ varies from $0$ to $j_{\rm ext}$, suggesting that this conjectured bound may translate into non-trivial upper and lower bounds on the Wilsonian parameters $d_k$. A quantitative assessment would require scanning the full multi-dimensional parameter space of the EFT coefficients and confronting the resulting shifts with observational precision; we leave such an analysis to future work.

There are several natural extensions of our results. Chief among these is the study of the EFT spectra of Kerr–Newman black holes. The QNM spectra of Kerr–Newman black holes were first disentangled in \cite{Dias:2015wqa} (and studied in detail in \cite{Dias:2022oqm}), and one could hope to adapt the technology developed here to that setting. The motivation for pursuing this direction stems from \cite{Horowitz:2024dch}, which argues that Kerr–Newman black holes violate EFT expectations at sufficiently small temperatures. It would therefore be important to determine whether an analogous phenomenon occurs for their EFT-corrected QNM spectra. This extension is technically non-trivial, since the analogue of the Hertz map is not presently known for Kerr–Newman backgrounds.

Finally, our results indicate an oscillatory approach to extremality controlled by $\log \tau_H$ in the grand-canonical ensemble. A near-horizon expansion—along the lines of \cite{TeukolskyPress1974,Detweiler1980,AnderssonGlampedakis2000,GlampedakisAndersson2001,Hod2008,YangEtAl2012,YangEtAl2013a,YangEtAl2013b,PhysRevD.93.044033}—would clarify whether these oscillations persist to arbitrarily small temperatures or instead characterise an intermediate regime. The approximately periodic modulation in $\log \tau_H$ is suggestive of discrete scale invariance, as arises in systems with complex scaling dimensions near quantum critical points. Since the EFT-deformed near-horizon extremal Kerr geometry exhibits an SO$(2,1)$ symmetry \cite{Horowitz:2023xyl}, such an interpretation is natural; nevertheless, this connection should presently be regarded as conjectural. An analytic near-horizon treatment would be required to establish whether the oscillatory behaviour reflects a genuine universal feature of the extremal limit.

{\it Acknowledgments.}
We are grateful to  \'Oscar~Dias, Maciej~Kolanowski, Harvey~Reall, and Pablo~Cano for their comments on an earlier draft. WLB was supported by an STFC studentship ST/Y509127/1. The work of JES was partially supported by STFC consolidated grant ST/X000664/1 and by Hughes Hall College.

\bibliographystyle{apsrev4-2}
\bibliography{prl}

\onecolumngrid
\appendix* 
\section{Supplemental Material}
\subsection{Equations of motion}
We work with the action Eq. (\ref{eq:GREFTB}) and assume that the characteristic length scale of the curvature of the system being studied is $\gg 1/\Lambda$, where $\Lambda$ is the energy scale of new physics, so that we can treat the Wilsonian coefficients $d_k$ as being infinitesimally small. From this action, we can then derive the equations of motion
\begin{equation}\label{eq: EFT Einstein eqn}
E_{ab} \equiv  R_{ab} - \frac{1}{2} g_{ab} R -\sum_{k=1}^5 d_k T^{(k)}_{ab}=0\, ,
\end{equation}
where
\be
\begin{aligned}
   \kappa^{-2}_{4D} \,T_{ab}^{(1)} &= -3 R_{acde} R_{b\phantom{c}fg}^{\phantom{b}c} R^{defg} +\frac{1}{2} g_{ab} R_{cd}^{\phantom{cd}ef}R_{ef}^{\phantom{ef}gh}R_{gh}^{\phantom{gh}cd} -6 
    \nabla^c \nabla^d\left(R_{(a|cef} R_{|b)d}^{\phantom{|b)d}ef}\right) \\
    \kappa^{-2}_{4D}\,T_{ab}^{(2)} &= g_{ab} R_{ab}^{\phantom{ab}cd} R_{cd}^{\phantom{cd}ef} {\tilde R}_{ef}^{\phantom{ef}ab} -5 R_{ajcd} R_{b\phantom{j}ef}^{\phantom{b}j} \tilde R^{cdef} - 6 \nabla_c \nabla_d \left(R_{(a| \phantom{c}ef}^{\phantom{(a|}c} \tilde R_{|b)}^{\phantom{|b)}def}\right)\\
    \kappa^{-4}_{4D}\,T_{ab}^{(3)}&= -\frac{1}{2}g_{ab} \mathcal C^2 -8 R_{a\phantom{c}b}^{\phantom{a}c\phantom{b}d}\nabla_c \nabla_d \mathcal C\\
    \kappa^{-4}_{4D}\,T_{ab}^{(4)}&= -\frac{1}{2}g_{ab} \tilde {\mathcal C}^2 -8 \tilde R_{a\phantom{c}b}^{\phantom{a}c\phantom{b}d}\nabla_c \nabla_d \tilde{\mathcal C}  \\
    \kappa^{-4}_{4D}\,T_{ab}^{(5)} &=  - \frac{1}{2} g_{ab} \mathcal C \tilde {\mathcal C}-4  R_{a\phantom{c}b}^{\phantom{a}c\phantom{b}d}\nabla_c \nabla_d \tilde{\mathcal C}-4 \tilde R_{a\phantom{c}b}^{\phantom{a}c\phantom{b}d}\nabla_c \nabla_d {\mathcal C}\, ,
    \end{aligned}
\ee
are the effective stress-energy tensors for this theory. Note that our infinitesimal treatment of the $d_k$s means that we have substituted $R_{ab}=0$ into our construction of the effective stress-energy tensors, greatly simplifying their form. In what follows this infinitesimal treatment also mean that each $d_k$ can be considered separately, so from now on we will drop the $k$ subscript to work with a generic EFT perturbation, with corresponding parameter $d$. 

\subsection{EFT corrected Kerr in the NP formalism}
The study of Kerr QNMs is most tractable in the Newman-Penrose (NP) formalism \cite{Newman:1961qr}, since the Kerr metric is algebraically special of type-D. As such, we will approach the problem of studying perturbations to EFT Corrected Kerr spacetime Eq. (\ref{eq:EFTKmetricmain}) in a similar manner. We take the following vector fields as our NP tetrad:
\begin{equation} \label{eq:EFTtetrad}
\begin{aligned}
\ell_{\rm EFTK} &=\frac{\Sigma(r,x)}{r^2\sqrt{F_1(r,x)}[a_{\rm K}^2 + r^2 - (1-x^2)F_4(r,x)^2]}\left[(r^2 + a_{\rm K}^2) \frac{\partial}{\partial t} + F_4(r,x)\frac{\partial}{\partial \phi}\right]+ \frac{\Delta_{\rm K}(r)}{r^2\sqrt{F_2(r,x)}}\frac{\partial}{\partial r}  \\
n_{\rm EFTK} &=\frac{r^2}{2\Delta_{\rm K}(r)\sqrt{F_1(r,x)}[a_{\rm K}^2 + r^2 - (1-x^2)F_4(r,x)^2]}\left[(r^2 + a_{\rm K}^2) \frac{\partial}{\partial t} + F_4(r,x)\frac{\partial}{\partial \phi}\right]- \frac{r^2}{2 \Sigma(r,x)\sqrt{ F_2(r,x)}}\frac{\partial}{\partial r}  \\
m_{\rm EFTK} &=\frac{\sqrt{1-x^2}}{(r+{\rm i} a_{\rm K} x)} \left\{ \frac{i\Sigma(r,x)}{\sqrt{2F_3(r,x)} [a_{\rm K}^2 + r^2 - (1-x^2) F_4(r,x)^2]} \left[ F_4(r,x) \frac{\partial}{\partial t} +  \frac{1}{1-x^2} \frac{\partial}{\partial \phi} \right]- \frac{1}{\sqrt{ F_2(r,x)}} \frac{\partial}{\partial x}\,\right\} ,
\end{aligned}
\end{equation}
which reduces to a boosted version of the NP tetrad of Kinnersley \cite{Kinnersley:1969zza} when we set $d=0$. This tetrad is particularly convenient as it is regular across the future event horizon, and $\ell_{\text{EFTK}}, n_{\text{EFTK}}$ reduce to repeated principal null vectors after reduction to Kerr.

From the metric Eq. (\ref{eq:EFTKmetricmain}), it will also be useful to read off the relevant thermodynamic variables
\begin{equation}
\begin{aligned}
    \Omega_H &= \frac{F_4(r_+, x)}{r_+^2+a_{\rm K}^2}\\
    \varkappa &= \frac{r_+ - M_{\rm K}}{(r_+^2 + a_{\rm K}^2 x^2)} \sqrt{\frac{F_1(r_+,x)}{F_2(r_+,x)}} \left[1- (1-x^2)  \Omega_H^2 (r_+^2 +a_{\rm K}^2) \right]\\
    E &= \frac{8\pi}{\kappa^2_{4D}}\left[M_K + \frac{1}{2} \lim_{r \to \infty} r^2 \frac{\partial}{\partial r} F_1(r,x)\right] \\
    j &= \frac{8\pi}{\kappa^{2}_{4D}M_{\rm K}}\lim_{r \to \infty} F_4(r,x),
\end{aligned}    
\end{equation}
obtained by smoothly continuing across the future event horizon and by performing an asymptotic expansion at future null infinity. Although these apparently depend on the polar coordinate $x$, in fact the equations of motion fix these to be independent of $x$, as would be expected from the first law of black hole thermodynamics and asymptotic expansions at future null infinity.

\subsection{The universal Teukolsky equation}
To study QNMs in a \emph{generic} spacetime which is perturbatively far from Kerr, we seek to generalise the Teukolsky equation \cite{Teukolsky:1973ha} describing the dynamics of certain components of the Weyl tensor in type-D spacetimes. This will provide us with a universal Teukolsky equation, as constructed in \cite{Campanelli:1998jv,Cano:2023univTeuk}, whose derivation we repeat here. By manipulating the NP field equations
\begin{equation}
\begin{aligned}
(D-\rho - \bar \rho - 3 \epsilon + \bar \epsilon) \sigma - (\delta - \tau + \bar \pi - \bar \alpha - 3 \beta) \kappa -\Psi_0 &=0 \\
(\Delta + \mu + \bar \mu + 3 \gamma - \bar \gamma) \lambda - (\bar\delta + 3 \alpha + \bar \beta + \pi - \bar \tau ) \nu +\Psi_4 &=0  \, ,
\end{aligned}
\end{equation}
and the Bianchi identities
\begin{equation}
\begin{aligned}
 ( \bar \delta + \pi - 4\alpha)\Psi_0 - 3 \kappa \Psi_2 &=  (D-4 \rho -2 \epsilon) \Psi_1 + (\delta +\bar \pi - 2 \bar \alpha - 2 \beta) \Phi_{00}- (D - 2\bar \rho -2 \epsilon)\Phi_{01}    + 2 \sigma \Phi_{10} - 2 \kappa \Phi_{11} - \bar \kappa \Phi_{02}\\
  (\Delta- 4\gamma + \mu) \Psi_0 -3 \sigma \Psi_2  &=  ( \delta - 4 \tau -2 \beta)\Psi_1 + (\delta+ 2 \bar \pi -2 \beta)\Phi_{01}  -( D - \bar \rho - 2 \epsilon + 2 \bar \epsilon)\Phi_{02}+ 2 \sigma \Phi_{11} - 2 \kappa \Phi_{12}   - \bar \lambda \Phi_{00}\\
  ( D-\rho + 4 \epsilon) \Psi_4 + 3 \lambda \Psi_2   &= (\bar \delta  + 2\alpha + 4 \pi) \Psi_3  - (\Delta + 2 \gamma- 2 \bar \gamma + \bar \mu) \Phi_{20}+ (\bar \delta - 2 \bar \tau +2 \alpha) \Phi_{21} + \bar \sigma \Phi_{22}  +   2 \nu \Phi_{10} - 2 \lambda \Phi_{11}\\
 (\delta +4 \beta - \tau)\Psi_4  + 3 \nu \Psi_2 &= (\Delta +2 \gamma + 4 \mu) \Psi_3   -(\Delta + 2 \gamma + 2 \bar \mu )\Phi_{21}+ (\bar \delta - \bar \tau + 2 \bar \beta + 2 \alpha) \Phi_{22} + 2 \nu \Phi_{11} + \bar \nu \Phi_{20} - 2 \lambda \Phi_{12} \, ,
\end{aligned}
\end{equation}
one can obtain the universal Teukolsky equations \cite{Campanelli:1998jv,Cano:2023univTeuk}
\begin{equation}\label{eq: Univ Teuk}
\begin{aligned}
    \mathcal T_+ \Psi_0 = \mathcal F_+,\qquad
    \mathcal T_- \Psi_4 = \mathcal F_- \, ,
\end{aligned}
\end{equation}
where $\Psi_{0,4}$ are Weyl scalars, and $\mathcal T_\pm, \mathcal F_\pm$ are given by
\begin{equation}
\begin{aligned}
\mathcal T_+ &=\Big(D-\rho - \bar \rho - 3 \epsilon + \bar \epsilon- \frac{1}{\Psi_2}D \Psi_2\Big)\Big(\Delta- 4\gamma + \mu\Big) 
- \Big(\delta - \tau + \bar \pi - \bar \alpha - 3 \beta - \frac{1}{\Psi_2} \delta \Psi_2\Big)\Big( \bar \delta + \pi - 4\alpha\Big) - 3 \Psi_2 \\
\mathcal T_- &=\Big(\Delta + \mu + \bar \mu + 3 \gamma - \bar \gamma- \frac{1}{\Psi_2}\Delta \Psi_2\Big)\Big( D-\rho + 4 \epsilon\Big)  - \left(\bar\delta + 3 \alpha + \bar \beta + \pi - \bar \tau- \frac{1}{\Psi_2}\bar \delta \Psi_2\right)\Big( \delta +4 \beta - \tau\Big) - 3 \Psi_2 \\
\mathcal F_+ &=\Big(D-\rho - \bar \rho - 3 \epsilon + \bar \epsilon- \frac{1}{\Psi_2}D \Psi_2\Big)\Big[( \delta - 4 \tau -2 \beta)\Psi_1 + (\delta+ 2 \bar \pi -2 \beta)\Phi_{01}  
\\ &\hspace{7cm} -( D - \bar \rho - 2 \epsilon + 2 \bar \epsilon)\Phi_{02} + 2 \sigma \Phi_{11} - 2 \kappa \Phi_{12}   - \bar \lambda \Phi_{00}\Big] \\
&\quad - \left(\delta - \tau + \bar \pi - \bar \alpha - 3 \beta - \frac{1}{\Psi_2} \delta \Psi_2\right)\Big[ (D-4 \rho -2 \epsilon) \Psi_1 + (\delta +\bar \pi - 2 \bar \alpha - 2 \beta) \Phi_{00}
\\ &\hspace{7cm}- (D - 2\bar \rho -2 \epsilon)\Phi_{01}   + 2 \sigma \Phi_{10} - 2 \kappa \Phi_{11} - \bar \kappa \Phi_{02}\Big] \\
\mathcal F_- &= \left(\Delta + \mu + \bar \mu + 3 \gamma - \bar \gamma- \frac{1}{\Psi_2}\Delta \Psi_2\right)\Big[ (\bar \delta  + 2\alpha + 4 \pi) \Psi_3  - (\Delta + 2 \gamma- 2 \bar \gamma + \bar \mu) \Phi_{20}\\ &\hspace{7cm}+ (\bar \delta - 2 \bar \tau +2 \alpha) \Phi_{21} + \bar \sigma \Phi_{22}  +   2 \nu \Phi_{10} - 2 \lambda \Phi_{11}\Big] \\
&\quad - \left(\bar\delta + 3 \alpha + \bar \beta + \pi - \bar \tau- \frac{1}{\Psi_2}\bar \delta \Psi_2\right)\Big[ (\Delta +2 \gamma + 4 \mu) \Psi_3   -(\Delta + 2 \gamma + 2 \bar \mu )\Phi_{21}\\ &\hspace{7cm}+ (\bar \delta - \bar \tau + 2 \bar \beta + 2 \alpha) \Phi_{22} + 2 \nu \Phi_{11} + \bar \nu \Phi_{20} - 2 \lambda \Phi_{12}\Big]\, .
\end{aligned}
\end{equation}
We now wish to study small perturbations to NP quantities which will then give rise to wave equations satisfied by the perturbations to the Weyl scalars $\Psi_{0,4}$. A convenient notation for expanding perturbations to NP quantities is as follows: For a general field $\psi$, we make an expansion
\begin{equation} \label{eq: superscript convention}
    \psi = \psi^{(0)} +  b \,\psi^{(b)} +d \,\kappa^{-2}_{4D} \, \psi^{(d)} + b \,d \, \kappa^{-2}_{4D} \,\psi^{(bd)} 
\end{equation}
where $\psi^{(0)}$ is the field as in the Kerr spacetime, $d$ is our EFT coefficient as in Eq. (\ref{eq:GREFTB}), and $b$ is a parameter corresponding to the perturbatively small amplitude of dynamical perturbations. We have included factors of $\kappa_{4D}$ to ensure that $\psi^{(0,b,d,bd)}$ all have the same dimension. In practice, calculations are most convenient working in units where $\kappa_{4D} = 1$, with factors of $\kappa_{4D}$ restored at the end. For a general global (e.g. thermodynamic) quantity $A$, we need only correct to $\mathcal O(d)$ and so expand as $ A = A^K + d \,\kappa^{-2}_{4D}\, \delta A$, where $A^K$ is the value in the $d=0$ theory for Kerr, and $\delta A$ is the correction. 

Recall that the Kerr spacetime is vacuum and algebraically special of type-D, so that the following NP quantities vanish
\begin{equation}
\begin{aligned}
    \Phi_{ij}^{(0, b)} &= 0\qquad  \text{(Vacuum)}\\
    \Psi_{0,1,3,4}^{(0)} &= 0\qquad  \text{(Type-D)}\\
    \kappa^{(0)} =\sigma^{(0)}=\nu^{(0)}=\lambda^{(0)} &= 0 \qquad\text{(Vacuum type-D)} \,,
\end{aligned}
\end{equation}
and that in vacuum type-D spacetimes we have
\begin{equation}
    D \Psi_2 = 3 \rho \Psi_2,\qquad \Delta \Psi_2 = - 3 \mu \Psi_2 , \qquad \delta \Psi_2 = 3 \tau \Psi_2,\qquad \overline{\delta} \Psi_2 = - 3 \pi \Psi_2\, ,
\end{equation}
along with the identities
\begin{equation}\label{eq: Kerr commutator identity}
\begin{aligned}
        \left(D-4\rho - \bar \rho - 3 \epsilon + \bar \epsilon\right)( \delta - 4 \tau -2 \beta) - \left(\delta - 4\tau + \bar \pi - \bar \alpha - 3 \beta \right) (D-4 \rho -2 \epsilon) &= 0\\
       \left(\Delta +4 \mu + \bar \mu + 3 \gamma - \bar \gamma\right) (\bar \delta  + 2\alpha + 4 \pi) - \left(\bar\delta + 3 \alpha + \bar \beta + 4\pi - \bar \tau\right) (\Delta +2 \gamma + 4 \mu) &= 0 \, .
\end{aligned}
\end{equation}
This greatly simplifies the $\mathcal O(b)$ term of Eq. (\ref{eq: Univ Teuk}) to give the standard Teukolsky equations \cite{Teukolsky:1973ha} $\mathcal T_+^{(0)} \Psi_0^{(b)} = 0, \mathcal T_-^{(0)} \Psi_4^{(b)} = 0$, where
\begin{equation}\label{eq: Teukolsky Operators 0}
\begin{aligned}
\mathcal T_+^{(0)} &=\Big[ \left(D-4\rho - \bar \rho - 3 \epsilon + \bar \epsilon\right)\left(\Delta- 4\gamma + \mu\right) - \left(\delta - 4\tau + \bar \pi - \bar \alpha - 3 \beta \right)( \bar \delta + \pi - 4\alpha) - 3 \Psi_2 \Big]\Bigg|_{\text{Kerr}}\\
\mathcal T_-^{(0)} &=\Big[(\Delta +4 \mu + \bar \mu + 3 \gamma - \bar \gamma)( D-\rho + 4 \epsilon) - (\bar\delta + 3 \alpha + \bar \beta + 4 \pi - \bar \tau)( \delta +4 \beta - \tau) - 3 \Psi_2 \Big] \Bigg|_{\text{Kerr}} \, .
\end{aligned}
\end{equation}
These are now second order PDEs on the Kerr background, which can be solved with standard numerical methods. Remarkably, these PDEs separate into ODEs: in the tetrad Eq. (\ref{eq:EFTtetrad}), the Weyl scalars admit the separation
\begin{equation}
    \Psi^{(b)}_0 = e^{- {\rm i} \omega t + {\rm i} m \phi} R_0(r) S_{+2}(x),\qquad \Psi^{(b)}_4 = e^{- {\rm i} \omega t + {\rm i} m \phi} (r-{\rm i} a_{\rm K} x)^{-4} R_4(r) S_{-2}(x)
\end{equation}
where $S_{\pm 2}(x)$ are spin-weighted spheroidal harmonics. The functions $R_{0,4}(r)$ and $S_{\pm 2}(x)$ obey second order ODEs as listed in \cite{Teukolsky:1973ha} after $R_{0,4}(r)$ has been rescaled to account for the fact that our tetrad is in fact boosted by a radially dependent function from that of \cite{Kinnersley:1969zza}, which is the tetrad used by \cite{Teukolsky:1973ha}.

\subsection{Reconstructing metric perturbations}
Although solutions to the Teukolsky equation determine the QNM spectrum of the Kerr black hole, this is not enough to reconstruct metric perturbations. For this, there are two approaches which can be taken. The first and conceptually clearest is that of Chandrasekhar \cite{Chandrasekhar:1985kt}. In this, Chandrasekhar uses some of the gauge freedom in perturbing NP tetrads and in diffeomorphisms to set $\Psi_{1,2,3}^{(b)}=0$. He then takes solutions to the Teukolsky equations for $\Psi_{0,4}^{(b)}$ (with a relative normalisation between these as yet undetermined) and performs an intricate analysis of the NP equations to reconstruct the metric and determine the relative normalisation between these Weyl scalars. The alternative and more practical method which we will follow is to construct metric perturbations via the method of Hertz potentials \cite{Wald:1978vm,Cohen:1974cm,COHEN19755,Chrzanowski:1975wv}. In this method, a theorem due to Wald \cite{Wald:1978vm} allows the construction of metric perturbations from a solution to the adjoint Teukolsky equation, which is called a Hertz potential. The Teukolsky operators have adjoints 
\begin{equation}
\begin{aligned}
        \left(\mathcal T^{(0)}_+\right)^{\dagger} &= \Big[ \left(\Delta + 3 \gamma - \bar \gamma + \bar \mu\right)\left(D + 4 \epsilon + 3 \rho\right) - \left(\bar\delta - \bar\tau + \bar \beta + 3 \alpha \right)(\delta +3 \tau + 4 \beta) - 3 \Psi_2 \Big]  \Bigg|_{\text{Kerr}}\\
    \left(\mathcal T^{(0)}_-\right)^{\dagger} &= \Big[ \left(D - 3 \epsilon + \bar \epsilon - \bar \rho\right)\left(\Delta-4\gamma-3\mu\right) - \left(\delta -3 \beta - \bar \alpha + \bar \pi \right)(\bar \delta - 4 \alpha - 3 \pi) - 3 \Psi_2 \Big]  \Bigg|_{\text{Kerr}} \, ,
    \end{aligned}
\end{equation}
and so we seek solutions to the equations 
\begin{equation}
    \left(\mathcal T^{(0)}_\pm\right)^{\dagger} \Psi^H_\pm = 0\, .
\end{equation}
In a vacuum type-D spacetime, the Teukolsky operators are related to their adjoints by $\Psi_2^{4/3} \left(\mathcal T_{\pm}^{(0)}\right)^\dagger \Psi_2^{-4/3} = \mathcal T_\mp^{(0)}$, so we can in fact construct Hertz potentials just from solutions to the Teukolsky equation.

To compute $\mathcal O(b)$ perturbations to NP quantities, we reconstruct perturbations to the metric from solutions to the (adjoint) Teukolsky equation using the Hertz reconstruction map \cite{Wald:1978vm,Cohen:1974cm,COHEN19755,Chrzanowski:1975wv}, which gives
\begin{equation}
\begin{aligned}
h_{ab}^{+}&= {\biggl \{} \ell_{(a}m_{b)}
\left[ (D+3\epsilon+\overline{\epsilon}-\rho+\overline{\rho})
(\delta+4\beta+3\tau)
 +(\delta+3\beta-\overline{\alpha}-\tau-\overline{\pi})(D+4\epsilon+3\rho)
 \right]   \\
 & \quad-\ell_a \ell_b (\delta+3\beta+\overline{\alpha}-\tau)
(\delta+4\beta+3\tau) -m_a m_b
(D+3\epsilon-\overline{\epsilon}-\rho) (D+4\epsilon+3\rho) {\biggl
\}} \Bigg |_{\text{Kerr}}\Psi^H_+\\
&\hspace{1.5cm}+{\rm c.c.}\,, \\
h_{ab}^{-}&= {\biggl \{}
n_{(a}\overline{m}_{b)} \left[
(\overline{\delta}+\overline{\beta}-3\alpha+\overline{\tau}+\pi)
(\Delta-4\gamma -3\mu)
 +(\Delta-3\gamma-\overline{\gamma}+\mu-\overline{\mu})(\overline{\delta}-4\alpha-3\pi)
 \right]  \\
 &\quad -n_a n_b (\overline{\delta}-\overline{\beta}-3\alpha+\pi)
(\overline{\delta}-4\alpha -3\pi) -\overline{m}_a \overline{m}_b
(\Delta-3\gamma+\overline{\gamma}+\mu) (\Delta-4\gamma-3\mu) {\biggl
\}}\Bigg|_{\text{Kerr}}\Psi^H_-\\
&\hspace{1.5cm}+{\rm c.c.}\,. 
\end{aligned}
\end{equation}
As discussed in the main text, this would naively seem to double the degrees of freedom in the background metric perturbations. It is then straightforward to read off the perturbations to the NP tetrad and from there compute all perturbations to NP quantities via standard methods. In particular, we apply NP boosts and rotations to fix a gauge where perturbations to the NP tetrad are in the form 
\begin{equation}
\begin{aligned}
    l^{(b)} &= p_1\, n^{(0)} \\
    n^{(b)} &= p_2\, l^{(0)} + p_3\, n^{(0)}  \\
    m^{(b)} &= p_4\, l^{(0)}+ p_5\, n^{(0)} +p_6\, m^{(0)} +p_7 \,\bar m^{(0)}
\end{aligned}    
\end{equation}
where $p_{1,2,3,6} \in \mathbb R$ and $p_{4,5,7} \in \mathbb C$ are some functions related to the components of the metric perturbation. In this gauge we can read off these coefficients from $h_{ab}$ to be
\begin{equation}
\begin{aligned}
p_1 &= \frac{1}{2} h_{ab}\left( \ell^{(0)}\right)^a \left(\ell^{(0)}\right)^b = -\frac{1}{2}  (\overline{\delta}-\overline{\beta}-3\alpha+\pi)
(\overline{\delta}-4\alpha -3\pi)\Big|_{\text{Kerr}}\, \Psi_-^H + \text{c.c.}\\
p_2 &= \frac{1}{2} h_{ab}\left( n^{(0)}\right)^a \left(n^{(0)}\right)^b =  -\frac{1}{2}(\delta+3\beta+\overline{\alpha}-\tau)
(\delta+4\beta+3\tau)\Big|_{\text{Kerr}}\,\Psi_+^H+ \text{c.c.}\\
p_3 &=  h_{ab}\left( \ell^{(0)}\right)^a \left(n^{(0)}\right)^b = 0\\
p_4 &=  h_{ab}\left( n^{(0)}\right)^a \left(m^{(0)}\right)^b
 =-  \frac{1}{2}\left[ (D+3\overline{\epsilon}+\epsilon-\overline{\rho}+\rho)
(\overline\delta+4\overline\beta+3\overline\tau)
 +(\overline\delta+3\overline\beta-\alpha-\overline\tau-\pi)(D+4\overline\epsilon+3\overline\rho)
 \right] \,\overline \Psi_+^H\\
p_5 &=  h_{ab}\left( l^{(0)}\right)^a \left(m^{(0)}\right)^b= -   \frac{1}{2}\left[
(\overline{\delta}+\overline{\beta}-3\alpha+\overline{\tau}+\pi)
(\Delta-4\gamma -3\mu)
 +(\Delta-3\gamma-\overline{\gamma}+\mu-\overline{\mu})(\overline{\delta}-4\alpha-3\pi)
 \right]\Psi^H_-\\
p_6 &= -\frac{1}{2} h_{ab}\left( m^{(0)}\right)^a \left(\bar m^{(0)}\right)^b = 0\\
p_7 &= -\frac{1}{2} h_{ab}\left( m^{(0)}\right)^a \left(m^{(0)}\right)^b= \frac{1}{2}(\Delta-3\gamma+\overline{\gamma}+\mu) (\Delta-4\gamma-3\mu) \Psi_-^H  + \frac{1}{2}(D+3\overline\epsilon-{\epsilon}-\overline\rho) (D+4\overline\epsilon+3\overline\rho)\overline\Psi_+^H 
\end{aligned}
\end{equation}
from which we can evaluate all $\mathcal O(b)$ perturbations to NP quantities via standard tetrad methods. 

\subsection{Correcting QNM frequencies}
We now know all the $\mathcal O(b)$ perturbations to quantities appearing in the universal Teukolsky equation (\ref{eq: Univ Teuk}). Using the tetrad Eq. (\ref{eq:EFTtetrad}), we can compute the $\mathcal O(d)$ corrections to the Weyl scalars and spin coefficients in terms of the functions $f_I(r,x)$ as in Eq. (\ref{eq:F expansion}), which we can determine numerically using the methods of \cite{Horowitz:2024dch}. The $\mathcal O(d)$ and $\mathcal O(b d)$ corrections to the coefficients of the Ricci tensor, $\Phi_{ij}^{(d)}$ and $\Phi_{ij}^{(b d)}$, can be computed directly by plugging in the Kerr metric and its dynamical perturbations at $\mathcal O(b)$ into the effective energy-momentum tensor of Eq. (\ref{eq: EFT Einstein eqn}). Finally, we make an NP boost to set $\Psi_{1,4}^{(bd)} = 0$. We now know enough perturbations to NP quantities to derive an equation for the $\mathcal O(bd)$ corrections to the relevant Weyl scalars. We do this by expanding the universal Teukolsky equation (\ref{eq: Univ Teuk}) in powers of $b, d$ to give 
\be
\label{eq: EFT Teuk}
\begin{aligned}
    \mathcal T^{(0)}_{+ } \tilde \Psi_{0} &= d\, \kappa_{4D}^{-2} \,\mathcal G_+\equiv d \, \kappa_{4D}^{-2}(\mathcal F^{(bd)}_{+} -  \mathcal T^{(d)}_+ \Psi_{0}^{(b)} - \mathcal T^{(b)}_+ \Psi_{0}^{(d)})\\
    \mathcal T^{(0)}_{- } \tilde \Psi_{4} &= d \, \kappa_{4D}^{-2}\,\mathcal G_-\equiv d \, \kappa_{4D}^{-2}(\mathcal F^{(bd)}_{-} -  \mathcal T^{(d)}_- \Psi_{4}^{(b)} - \mathcal T^{(b)}_- \Psi_{4}^{(d)})\,.
\end{aligned}
\ee
where we have defined
\begin{equation}
    \tilde \Psi_{0,4} = \Psi_{0,4}^{(b)} + d\, \kappa^{-2}_{4D} \Psi_{0,4}^{(bd)}
\end{equation}
which we can think of as the dynamical part of the perturbations to these Weyl scalars. Note that $\mathcal G_{\pm}$ on the RHS of Eq. (\ref{eq: EFT Teuk}) are now functions we can completely calculate in a given QNM sector. 

\subsection{Boundary conditions}
Near $r = r_+$, one can define coordinates $(v,r,x,\chi)$ such that the metric in these coordinates can be smoothly continued across the event horizon. These behave as 
\begin{equation}
\begin{aligned}
    v &\sim t + \frac{1}{2 \varkappa} \log \left(\frac{r - r_+}{r_+}\right) \\
    \chi &\sim \phi + \frac{\Omega_H }{2 \varkappa} \log \left(\frac{r - r_+}{r_+}\right)\\
\end{aligned}
\end{equation}
We can similarly define coordinates $(u,r,x,\phi)$ near future null infinity to put the metric into Bondi-Sachs gauge, which near future null infinity behave as
\begin{equation}
    u \sim t -r - 2   M \log \left(\frac{r}{r_+}\right)\, ,
\end{equation}
where here $M = \frac{\kappa^2_{4D}}{8 \pi} E$ is the mass of this spacetime.
To place our equations on a compact domain, we define the coordinate $y = 1 - r_+/r$, where $r_+= M_K + \sqrt{M_K^2 - a^2_K}$ is the \emph{coordinate} horizon radius. Now $y= 0$ corresponds to the horizon and $y = 1$ corresponds to asymptotic infinity. 

In order to study QNMs, we want to factorise the dynamical part of the relevant Weyl scalars as
\begin{equation}\label{eq: weylfactorisation}
    \tilde \Psi_{0,4}(t,r,x,\phi) = e^{-{\rm i}  \omega t + {\rm i} m \phi}\, b_{0,4}(r,x) \, \hat Y_{0,4}(r,x)
\end{equation}
where $\hat Y_{0,4}$ is regular at the $ y = 0, 1$ boundaries and the $x = \pm 1$ boundaries. To achieve this, we have factored out some boundary `regulator' functions defined by
\begin{equation}
\begin{aligned}
    b_0(r,x) &= (1+x)^{|m-2|/2} (1-x)^{|m+2|/2} \left(1- \frac{r_+}{r}\right)^{-{\rm i} (\omega - \Omega_H m)/2\varkappa} \left(\frac{r}{r_+}\right)^{2   M  {\rm i}  \omega-5} e^{ {\rm i} \omega r}\\
    &= (1+x)^{|m-2|/2} (1-x)^{|m+2|/2} y^{-{\rm i} (\omega - \Omega_H m)/2\varkappa} (1-y)^{-2   M  {\rm i} \omega+5} e^{{\rm i} \omega r_+/(1-y)} \\
    b_4(r,x) &= (1+x)^{|m+2|/2} (1-x)^{|m-2|/2} \left(1- \frac{r_+}{r}\right)^{-{\rm i} (\omega - \Omega_H m)/2\varkappa} \left(\frac{r}{r_+}\right)^{2  M  {\rm i}  \omega-1} e^{ {\rm i} \omega r} \\
    &= (1+x)^{|m+2|/2} (1-x)^{|m-2|/2} y^{-{\rm i} (\omega - \Omega_H m)/2\varkappa} (1-y)^{-2   M  i \omega+1} e^{{\rm i}  \omega r_+/(1-y)}\, .
\end{aligned}
\end{equation}
To see that these are the correct boundary behaviours requires expanding the NP tetrad Eq. (\ref{eq:EFTtetrad}) in suitable coordinates across the horizon, near future null infinity, and at the poles $x = \pm 1$. As we have already seen, all $\mathcal O(b)$ quantities can be constructed by linear differential operators acting on Hertz potentials $\Psi_{\pm}^{H}$ or their complex conjugates, so the RHS of Eq. (\ref{eq: EFT Teuk}) must also be constructed as a linear differential operator acting on these functions. To study the QNMs of this problem, we also take $\Psi_\pm^H$ to have a time dependence factor $e^{-{\rm i}  \omega t}$ (or equivalently $e^{-{\rm i} \omega_K t}$ since any term involving the Hertz potentials has a $d$ prefactor). Using the relationship between the Teukolsky operators and their adjoints, we deduce that the Hertz potentials can be separated as in \cite{Teukolsky:1973ha} into radial and angular parts, explicitly:
\begin{equation}
\begin{aligned}
    \Psi_+^H(t,r,x,\phi) &= e^{- {\rm i} \omega t + {\rm i} m \phi} R_+^H(r) S_+^H(x) \\
    \Psi_-^H(t,r,x,\phi) &= e^{- {\rm i} \omega t + {\rm i} m \phi} (r-{\rm i} a_{\rm K} x)^4 R_-^H(r) S_-^H(x)
\end{aligned}
\end{equation}
Note that the relationship between the Teukolsky equation and its adjoint determines that $R_+^H(r)=R_4(r),R_-^H(r)=R_0(r), S_+^H(x) =S_{-2}(x), S_-^H(x) =S_{+2}(x)$, with the notation changed here for convenience. We can deduce the relevant boundary conditions for $R_\pm^H$ and $S_\pm^H$ by evaluating the Weyl scalars $\Psi_{0,4}^{(b)}$ generated by these Hertz potentials, which allows us to read off that the appropriate factorisation is
\begin{equation}
R^H_\pm(r) = b^H_{r,\pm}(r) Z^R_\pm(r)\,,\quad S^H_\pm(x) = b^H_{x,\pm}(x)Z^S_\pm(x)    \, ,
\end{equation}
where
\begin{equation}
\begin{aligned}
b^H_{r,+}(r) &= \left(1-\frac{r_+}{r}\right)^{-{\rm i}(\omega - m \Omega_H)/2\varkappa} \left( \frac{r}{r_+}\right)^{2 M {\rm i} \omega +3} e^{{\rm i} \omega r}\\
b^H_{r,-}(r) &= \left(1-\frac{r_+}{r}\right)^{-{\rm i}(\omega - m \Omega_H)/2\varkappa} \left( \frac{r}{r_+}\right)^{2 M {\rm i} \omega -5} e^{{\rm i} \omega r}\\
b^H_{x,+}(x) &=(1+x)^{|m+2|/2} (1-x)^{|m-2|/2} \\
b^H_{x,-}(x) &= (1+x)^{|m-2|/2} (1-x)^{|m+2|/2}\, .
\end{aligned}
\end{equation}
After these factorisations, the LHS of Eq. (\ref{eq: EFT Teuk}) has time dependence $e^{-{\rm i}  \omega t+ {\rm i} m \phi }$ whereas the RHS has an $e^{-{\rm i}  \omega t+ {\rm i} m \phi}$ term coming from terms constructed from $\Psi_\pm^H$ but also an $e^{{\rm i} \bar{  \omega} t- {\rm i} m \phi}$ term coming from terms constructed from $\bar{\Psi}_\pm^H$. By taking a suitably-weighted Fourier transform we can remove the term constructed from $\bar{\Psi}_H^\pm$, so we need only consider those constructed without any complex conjugation. After doing so we can drop all $t$ and $\phi$ dependence from Eq. (\ref{eq: EFT Teuk}) and expand out $ \omega = \omega + d\, \kappa^{-2}_{4D} \, \delta\omega$. 
This allows us to express the RHS of Eq. (\ref{eq: EFT Teuk}) solely in terms of $Z^{R,S}_\pm$. We expand out $\hat Y_{0,4}(y,x)$ as
\begin{equation}
\begin{aligned}
    \hat Y_0(y,x) &= \chi_0 \, Y_0(y,x) + d\,\kappa_{4D}^{-2}\,\delta Y_0(y,x) \\
    \hat Y_4(y,x) &= \chi_4 \,Y_4(y,x) + d\,\kappa_{4D}^{-2}\,\delta Y_4(y,x)
\end{aligned}
\end{equation}
where $Y_{0,4}$ are some known solutions to the transformed Teukolsky equations, and $\chi_{0,4}$ are arbitrary constants. After expanding everything out in Eq. (\ref{eq: EFT Teuk}), we find the following equations
\begin{equation}\label{eq: reg eqns}
\begin{aligned}
    {\mathfrak t}_{0, \omega} [Y_0] &= 0 \\
    {\mathfrak t}_{4, \omega} [Y_4] &= 0 \\
    {\mathfrak t}_{0, \omega} [\delta Y_0] &= \chi_0 \,{\mathfrak g}_{00}[Y_0] + \chi_4 \,{\mathfrak g}_{04}[Y_4] -\chi_0 \, \delta \omega \, (\partial_\omega {\mathfrak t_{0,\omega}})[ Y_0]    \\
    {\mathfrak t}_{4, \omega} [\delta Y_4] &= \chi_0 \,{\mathfrak g}_{40}[Y_0] + \chi_4 \,{\mathfrak g}_{44}[Y_4] -\chi_4 \, \delta \omega \, (\partial_\omega {\mathfrak t_{4,\omega}})[ Y_4]    
\end{aligned}
\end{equation}
where ${\mathfrak t}_{i, \omega}$ are the standard Teukolsky operators after the transformations of this section, and ${\mathfrak g}_{ij}$ are some differential operators, constructed from $f_I(r,x)$, which depend on our choice of EFT correction. We are free to redefine $\delta Y_{0,4} \mapsto \delta Y_{0,4} + \lambda_{0,4}\, Y_{0,4}$ for any constants $\lambda_{0,4}$, so to remove this ambiguity, we impose $Y_{0,4}(0, 0) = 0$, alongside regularity at the boundaries of $(y,x) \in [0,1] \times [-1,1]$. We are also free to rescale $(\chi_0, \chi_4) \mapsto \lambda (\chi_0, \chi_4)$ for $\lambda \in \mathbb C^*$. To account for this ambiguity, we check for solutions with $\chi_0 = 1$ and then solutions with $\chi_4 = 1$, which covers the entire projective space of choices $(\chi_0, \chi_4)$.


\subsection{Numerical methods}
\subsection{Background solution}

Our computation proceeds in two numerical stages. We first construct the EFT‑deformed Kerr background. Following \cite{Horowitz:2024dch} and using the numerical framework of \cite{Dias:2015nua}, we exploit the fact that among the six non‑trivial components of the Einstein equation,
\begin{equation}
\{E^{tt}\,,E^{t\phi}\,,E^{\phi\phi}\,,E^{rr}\,,E^{rx}\,,E^{xx}\}
\end{equation}
one can select a set of four ``dynamical'' equations,
\begin{equation}
\{E^{tt}\,,E^{t\phi}\,,E^{\phi\phi}\,,E^{rr}g_{rr}+E^{xx}g_{xx}\}
\end{equation}
whose solution ensures that the full system is satisfied, provided the boundary conditions on the functions $f_I^{(k)}(r,x)$ in Eq.~(\ref{eq:F expansion}) are imposed consistently. The remaining equations,
\begin{equation}
\xi^1 \equiv \frac{\sqrt{\Delta_{\rm K}(r)}\sqrt{1-x^2}}{2}\,\Sigma(r,x)\left(E^{rr}g_{rr}-E^{xx}g_{xx}\right)\quad \text{and}
\quad
\xi^2 \equiv (1-x^2)\,\Sigma(r,x)\,g_{xx}\,E^{rx}\,,
\label{eq:con}
\end{equation}
act as constraints, which we evaluate after solving the dynamical system (see details in \cite{Horowitz:2024dch} for why this is the case). Demonstrating that the constraint violations decrease as the numerical resolution is increased provides a stringent and independent check of convergence.

All dynamical equations take the schematic form
\begin{equation}
\widetilde{\Delta} f_I^{(k)} + Q_I^{(k)}(f_J, \widetilde{\nabla} f_P, r, x) = S_I^{(k)}(r,x)\,,
\end{equation}
where $Q_I^{(k)}$ is linear in the fields $f_J$ and their first derivatives, and where $\widetilde{\Delta}$ denotes the Laplacian on the two‑dimensional orbit space
\begin{equation}
\frac{\dd r^2}{\Delta_{\rm K}(r)} + \frac{\dd x^2}{1 - x^2}\,.
\end{equation}

To implement the numerics, we introduce a compact radial coordinate
\begin{equation}
r = \frac{r_+}{1 - y}\,,
\end{equation}
placing the horizon at $y=0$ and spatial infinity at $y=1$. We discretise the $(x,y)$ domain using Chebyshev–Gauss–Lobatto grids with $N_x$ and $N_y$ points along $x\in[-1,1]$ and $y\in[0,1]$. After discretisation, the equations reduce to the linear system
\begin{equation}
\mathbf{A} \vec{f} = \vec{S}\,,
\end{equation}
where $\vec{f}$ is a $4 (N_x+1)(N_y+1)$-dimensional vector collecting the values of the four functions $f_I^{(k)}$ at all collocation points. We solve this system using a multifrontal LU decomposition in extended precision.

Accurate computation of the asymptotic coefficients requires evaluating up to six derivatives of $f_I^{(k)}(x,y)$ near infinity. For most of parameter space we use $N_x = N_y = 100$.

\subsection{Quasinormal modes and perturbative shifts}

We next compute the perturbed QNMs and their EFT‑induced frequency shifts. As discussed in subsection {\bf Correcting QNM frequencies}, this reduces to solving the system
\begin{subequations}
\begin{equation}
\label{eq:regeqnsNa}
\begin{aligned}
    {\mathfrak t}_{0, \omega} [Y_0] & = 0 \\
    {\mathfrak t}_{4, \omega} [Y_4] & = 0
\end{aligned}
\end{equation}
\begin{equation}
\label{eq:regeqnsNb}
\begin{aligned}
    {\mathfrak t}_{0, \omega} [\delta Y_0] & = \chi_0 \,{\mathfrak g}_{00}[Y_0] + \chi_4 \,{\mathfrak g}_{04}[Y_4] -\chi_0 \, \delta \omega \, (\partial_\omega {\mathfrak t_{0,\omega}})[ Y_0]    \\
    {\mathfrak t}_{4, \omega} [\delta Y_4] & = \chi_0 \,{\mathfrak g}_{40}[Y_0] + \chi_4 \,{\mathfrak g}_{44}[Y_4] -\chi_4 \, \delta \omega \, (\partial_\omega {\mathfrak t_{4,\omega}})[ Y_4]
\end{aligned}
\end{equation}
\end{subequations}%
where ${\mathfrak t}_{i,\omega}$ are second‑order differential operators in $(x,y)$, and the sources ${\mathfrak g}_{ij}[Y_j]$ are linear in $Y_j$ and their first derivatives.

The two equations in (\ref{eq:regeqnsNa}) are recast as a standard quadratic eigenvalue problem (following \cite{Dias:2015nua}), which determines the Kerr QNM spectrum $(\omega, Y_0, Y_4)$. The equations in (\ref{eq:regeqnsNb}) then determine $\chi_4/\chi_0$ and the perturbative corrections $\delta\omega$, $\delta Y_0$, and $\delta Y_4$. We discretise this system on the same $(x,y)$ grid used for the background fields, reducing it to a solvability‑condition system characteristic of degenerate perturbation theory:
\begin{equation}
\begin{aligned}
&\mathbf{A}_0 \vec{\delta Y}_0 = \chi_0\,\vec{G}_{00} + \chi_4\,\vec{G}_{04} - \chi_0\,\delta\omega\,\vec{T}_0
\\
&\mathbf{A}_4 \vec{\delta Y}_4 = \chi_0\,\vec{G}_{40} + \chi_4\,\vec{G}_{44} - \chi_4\,\delta\omega\,\vec{T}_4
\end{aligned}
\end{equation}
where the matrices $\mathbf{A}_i$ and vectors $\vec{G}_{ij}, \vec{T}_i$ are known functions of the background solution.

\subsection{Left‑null vectors and bordered system}

To solve these systems, we require the left‑null vectors of $\mathbf{A}_i$,
\begin{equation}
\vec{z}_i^{\mathsf{T}}\, \mathbf{A}_i = 0\Rightarrow \mathbf{A}^{\mathsf{T}}_i\vec{z}_i=0\,,
\end{equation}
with no summation over $i$. Instead of directly computing the nullspace of $\mathbf{A}_i^{\mathsf{T}}$, we impose a normalisation condition using a bordering vector $\vec{Y}_i$ and solve the bordered system
\begin{equation}
\left[
\begin{array}{cc}
\mathbf{A}_i^{\mathsf{T}} & \vec{Y}_i
\\
\vec{Y}_i^{\mathsf{T}} & 0
\end{array}
\right]
\left[
\begin{array}{c}
\vec{z}_i\\
\lambda
\end{array}
\right]=
\left[
\begin{array}{c}
0
\\
1
\end{array}
\right]\,.
\end{equation}
A natural and numerically robust choice for $\vec{Y}_i$ is the right‑null vector of $\mathbf{A}_i$; in our problem this is precisely the discretised versions of the QNM eigenfunctions $\{Y_0,Y_4\}$. This choice enforces the biorthogonality condition $\vec{Y}_i^{T}\vec{z}_i=1$ and guarantees consistency of the augmentation. Because $\mathbf{A}_i \vec{Y}_i = 0$, the bordered system automatically yields $\lambda = 0$. This method is standard in singular linear solves and degenerate perturbation theory \cite{Govaerts1991}.

\subsection{Solvability conditions and frequency shift}

Once the left‑null vectors $\vec{z}_i$ are known, determining $\chi_0/\chi_4$ and $\delta\omega$ is immediate. Acting with $\vec{z}_0$ ($\vec{z}_4$) on the first (second) equation in (\ref{eq:regeqnsNb}) gives
\begin{equation}
\begin{aligned}
& 0= \chi_0\,\vec{z}_0^{\mathsf{T}}\cdot\vec{G}_{00} + \chi_4\,\vec{z}_0^{\mathsf{T}}\cdot\vec{G}_{04} - \chi_0\,\delta\omega\,\vec{z}_0^{\mathsf{T}}\cdot\vec{T}_0
\\
&0 = \chi_0\,\vec{z}_4^{\mathsf{T}}\cdot\vec{G}_{40} + \chi_4\,\vec{z}_4^{\mathsf{T}}\cdot\vec{G}_{44} - \chi_4\,\delta\omega\,\vec{z}_4^{\mathsf{T}}\cdot\vec{T}_4
\end{aligned}
\end{equation}
which can be trivially reduced to a quadratic equation for $\delta \omega$ (or alternatively, for $\chi_0/\chi_4$).

\section{Parity odd splitting}
The quasinormal boundary conditions do not define a useful adjoint differential operator. However, after Chebyshev collocation, the equations for determining the shifts take the discrete form
\begin{equation}
\begin{aligned}
&\mathbf{A}_0 \,\vec{\delta Y}_0 \;=\; \chi_0\,\vec{G}_{00} + \chi_4\,\vec{G}_{04} \;-\; \chi_0\,\delta\omega\,\vec{T}_0,\\
&\mathbf{A}_4 \,\vec{\delta Y}_4 \;=\; \chi_0\,\vec{G}_{40} + \chi_4\,\vec{G}_{44} \;-\; \chi_4\,\delta\omega\,\vec{T}_4,
\end{aligned}
\label{eq:discrete-system}
\end{equation}
and solvability is enforced purely algebraically by left null vectors of the transposed matrices. Let $\vec{z}_i \in \ker(\mathbf{A}_i^{\mathsf T})$ denote the left null vectors (one for each block $i\in\{0,4\}$). Then the Fredholm alternative reduces to the discrete orthogonality conditions
\begin{equation}
\vec{z}_i^{\mathsf T}\!\left(\chi_0\,\vec{G}_{i0} + \chi_4\,\vec{G}_{i4} - \chi_i\,\delta\omega\,\vec{T}_i\right)=0,
\qquad i\in\{0,4\}.
\label{eq:solvability}
\end{equation}

The discretised operators inherit the parity symmetry $\mathbf{P}\,\mathbf{A}_i=\mathbf{A}_i\,\mathbf{P}$ with $\mathbf{P}^2=\mathbf{I}$, so $\mathbf{A}_i$ is block‑diagonal in the even/odd subspaces. Consequently, the right null vector (the unperturbed mode) and the left null vector $\vec{z}_i$ can both be chosen with definite and \emph{matching} parity. By assumption, the sources $\vec{G}_{ii}$ and $\vec{G}_{i0}$ have \emph{opposite} parity to $\vec{z}_i$, hence
\begin{equation}
\vec{z}_i^{\mathsf T}\vec{G}_{ii}=0,
\qquad
\vec{z}_i^{\mathsf T}\vec{G}_{i0}=0,
\label{eq:parity-orthogonality}
\end{equation}
and only the frequency‑shift term survives in \eqref{eq:solvability}. Therefore,
\begin{equation}
\vec{z}_i^{\mathsf T}\vec{T}_i \;\neq\; 0
\quad\Longrightarrow\quad
\chi_i\,\delta\omega\,\vec{z}_i^{\mathsf T}\vec{T}_i \;=\; 0,
\qquad i\in\{0,4\}.
\label{eq:single-block}
\end{equation}
The two parity‑related blocks yield a homogeneous $2\times 2$ system for $(\delta\omega,\chi_0/\chi_4)$; its compatibility condition is odd under the $\mathbb{Z}_2$ parity, so the frequency shifts appear in parity‑related pairs,
\begin{equation}
\delta\omega \;=\; \pm\,\delta\omega_{\rm phys}.
\label{eq:pm-branches}
\end{equation}
 
\subsection{Convergence plots}
Because the discretisation employs Chebyshev–Gauss–Lobatto collocation in both $x$ and $y$, the error should decrease exponentially with the number of collocation points $N=N_x=N_y$.

\begin{figure}
    \includegraphics[width=0.45\linewidth]{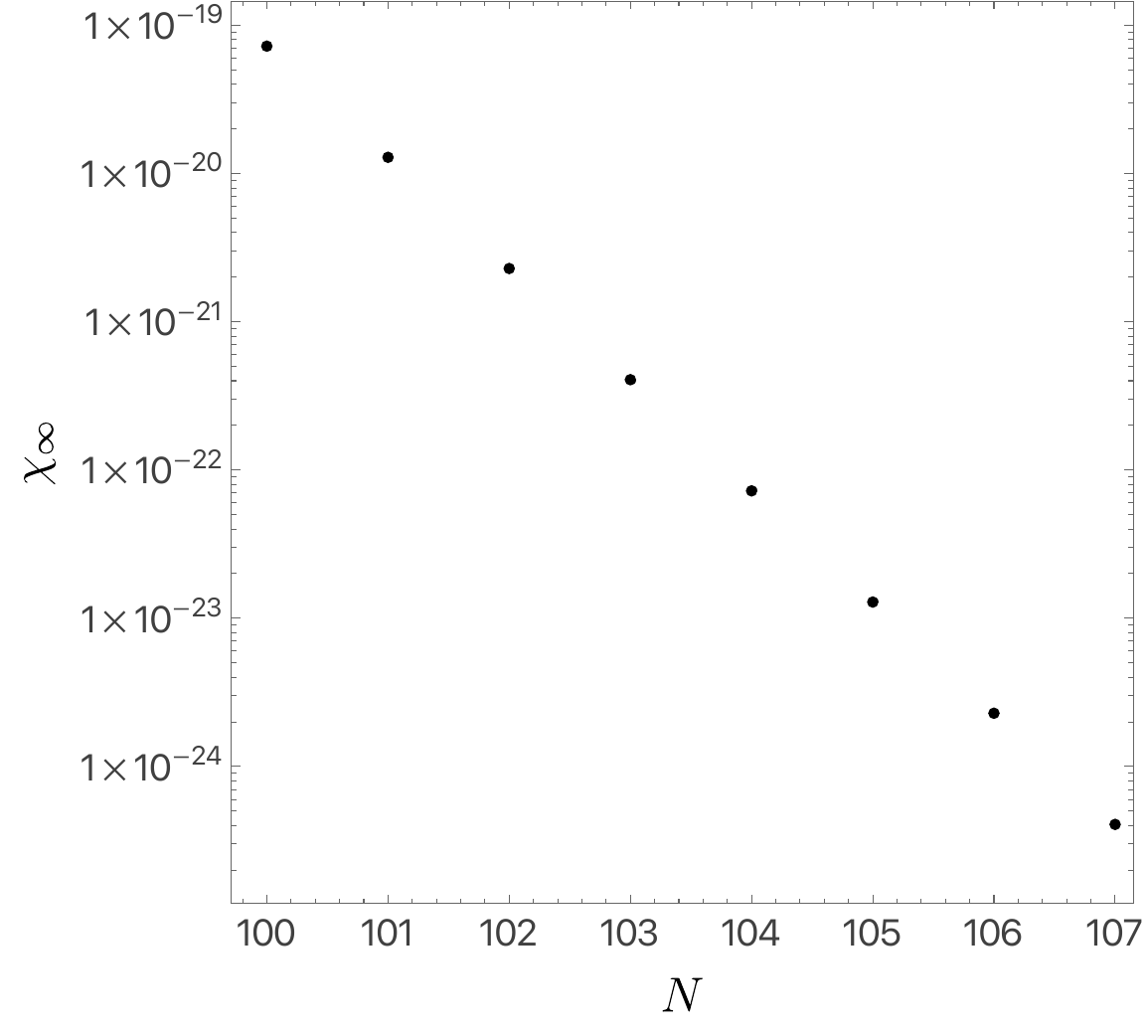}
    \caption{\label{fig:con1}Maximum constraint violation $\chi_{\infty}$ as a function of the number of collocation points $N$ for the EFT deformation proportional to $d_1$, evaluated at the near‑extremal spin $j/j_{\rm ext}=0.998686$. The clear exponential decay demonstrates the expected spectral convergence of the Chebyshev–Gauss–Lobatto scheme.}
\end{figure}

Let $\vec{\xi}^I_{N}$ denote the discretised constraints \eqref{eq:con} on a grid with $N = N_x = N_y$ collocation points. In Fig.~\ref{fig:con1} we show a log plot of
\begin{equation}
\chi_{\infty} \equiv \max\{\lVert \vec{\xi}^1 \rVert_{\infty},\, \lVert \vec{\xi}^2 \rVert_{\infty}\}
\end{equation}
as a function of $N$ for the EFT deformation proportional to $d_1$. The plot is produced for the near-extremal value $j/j_{\rm ext} = 0.998686$, where gradients are largest. The observed exponential convergence with $N$ is precisely the expected behaviour for Chebyshev–Gauss–Lobatto spectral collocation methods.

\begin{figure}
    \includegraphics[width=0.45\linewidth]{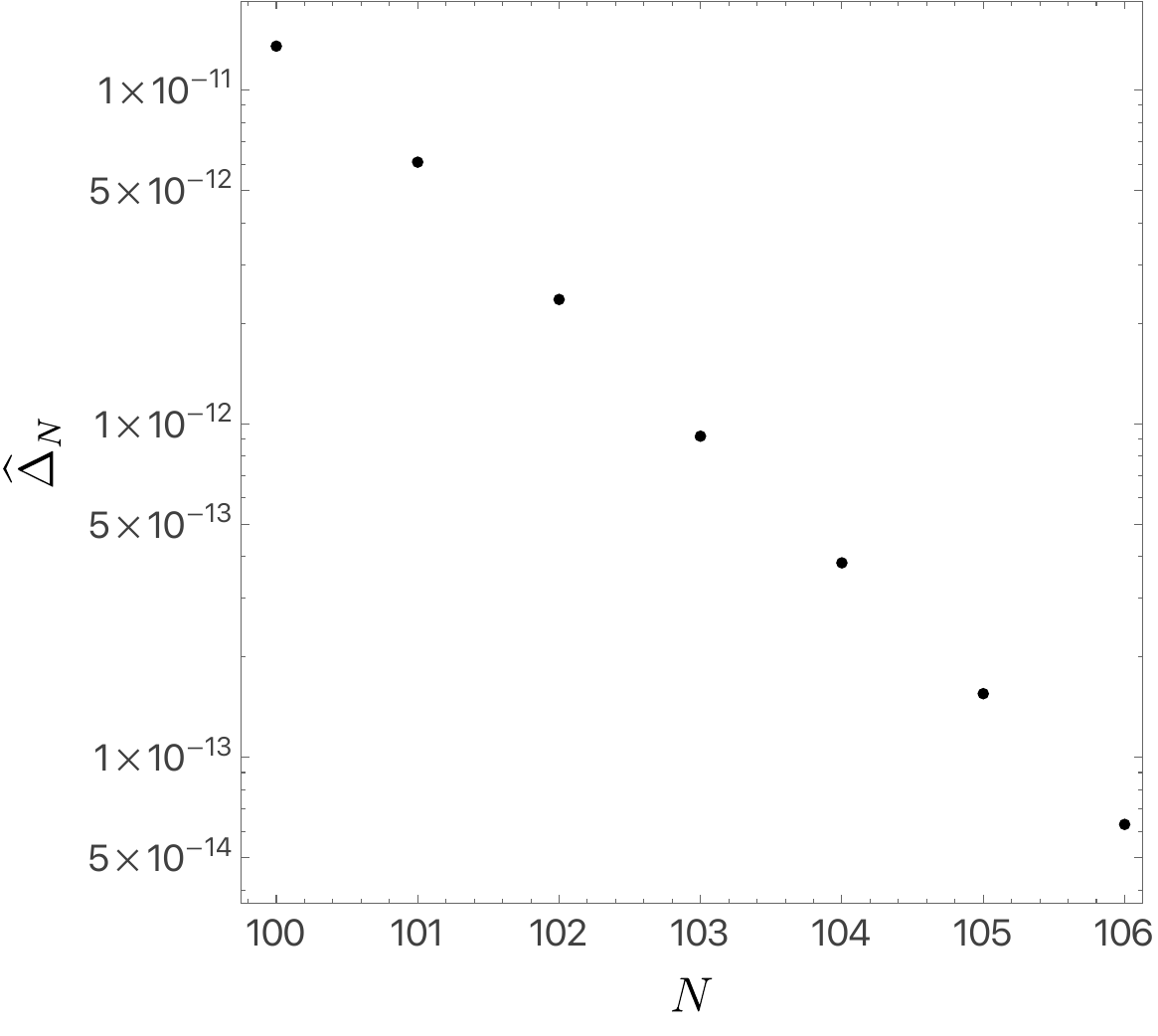}
    \caption{\label{fig:con2}Convergence diagnostic $\widehat{\Delta}_N$ for the QNM frequency shift $\delta\omega_{220}$ in the $d_1$ deformation at near‑extremal spin $j/j_{\rm ext}=0.998686$. The exponential decay with $N$ reflects the spectral accuracy of the Chebyshev–Gauss–Lobatto scheme.}
\end{figure}

To quantify the convergence of the numerical scheme used in the second stage of our calculation (where we compute the QNM frequency shifts), we consider
\begin{equation}
\widehat{\Delta}_N=\left|1-\frac{\delta \omega_{220}^N}{\delta \omega_{220}^{N+1}}\right|\,,
\end{equation}
where $\delta\omega_{220}^N$ is the QNM frequency shift obtained using a grid with $N$ collocation points. Since the large‑spin regime is the most demanding part of the parameter space, we test convergence at $j/j_{\rm ext}=0.998686$. Rather than presenting results for all EFT coefficients, we focus here on the deformation controlled by $d_1$. In Fig.~\ref{fig:con2} we show a log plot of $\widehat{\Delta}_N$, which again exhibits the expected exponential convergence with $N$.

\subsection{EFT deformations of the remaining QNMs associated with $d_i$ for $i=2,\ldots,5$}
\begin{figure*}[!t]
    \includegraphics[width=0.7\linewidth]{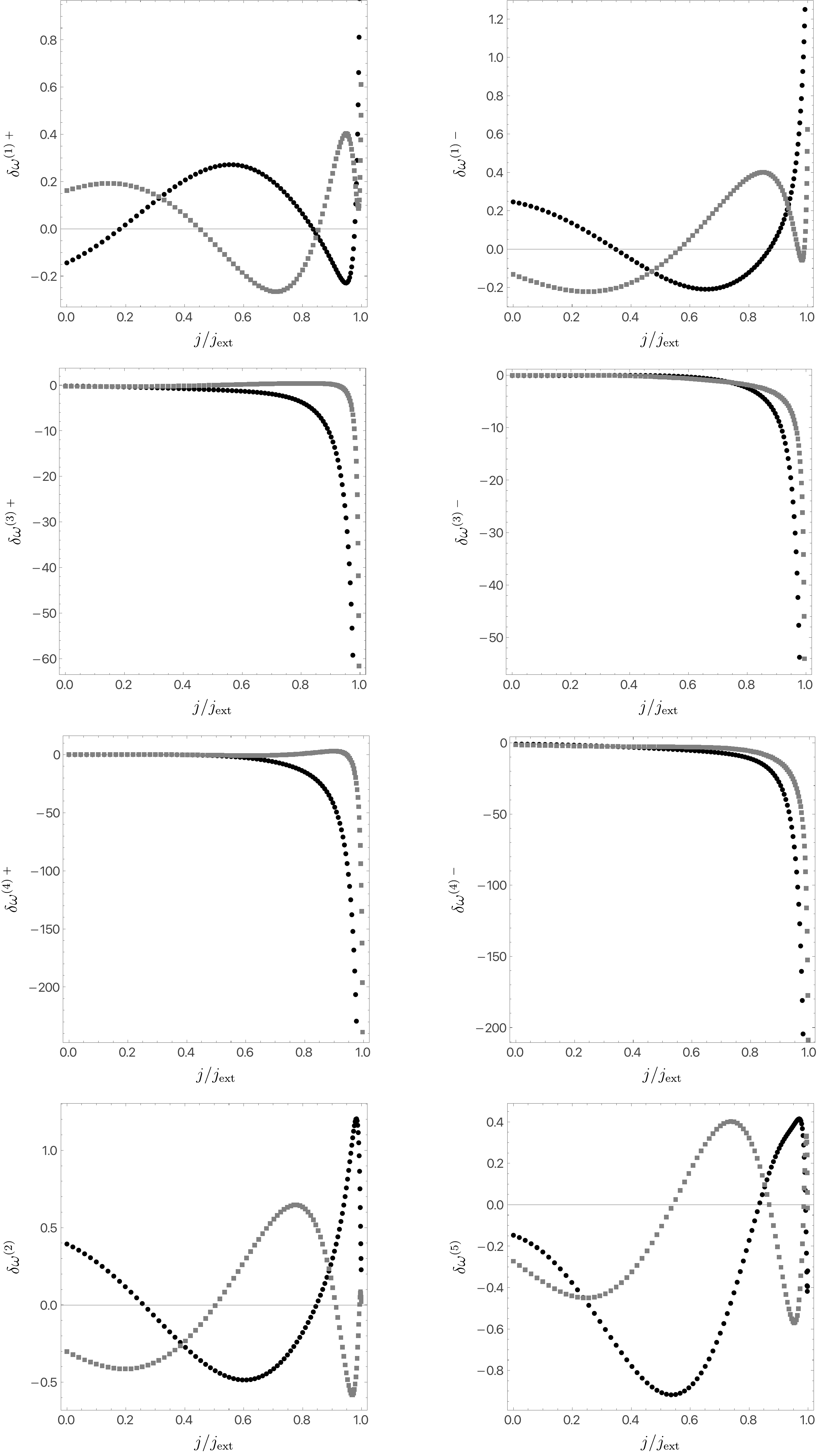}
    \caption{\label{fig:data1A}
$M_{\rm K}\,\mathrm{Re}\,\delta\omega^{(k)\pm}_{220}$ (black disks) and 
$M_{\rm K}\,\mathrm{Im}\,\delta\omega^{(k)\pm}_{220}$ (grey squares) as functions of 
$j/j_{\rm ext}$. The different sectors are indicated on the right of each plot. The data are shown in the microcanonical ensemble, where the EFT black holes and 
the Kerr black holes are compared at fixed energy and angular momentum.}
\end{figure*}

\begin{figure*}[!t]
    \includegraphics[width=0.7\linewidth]{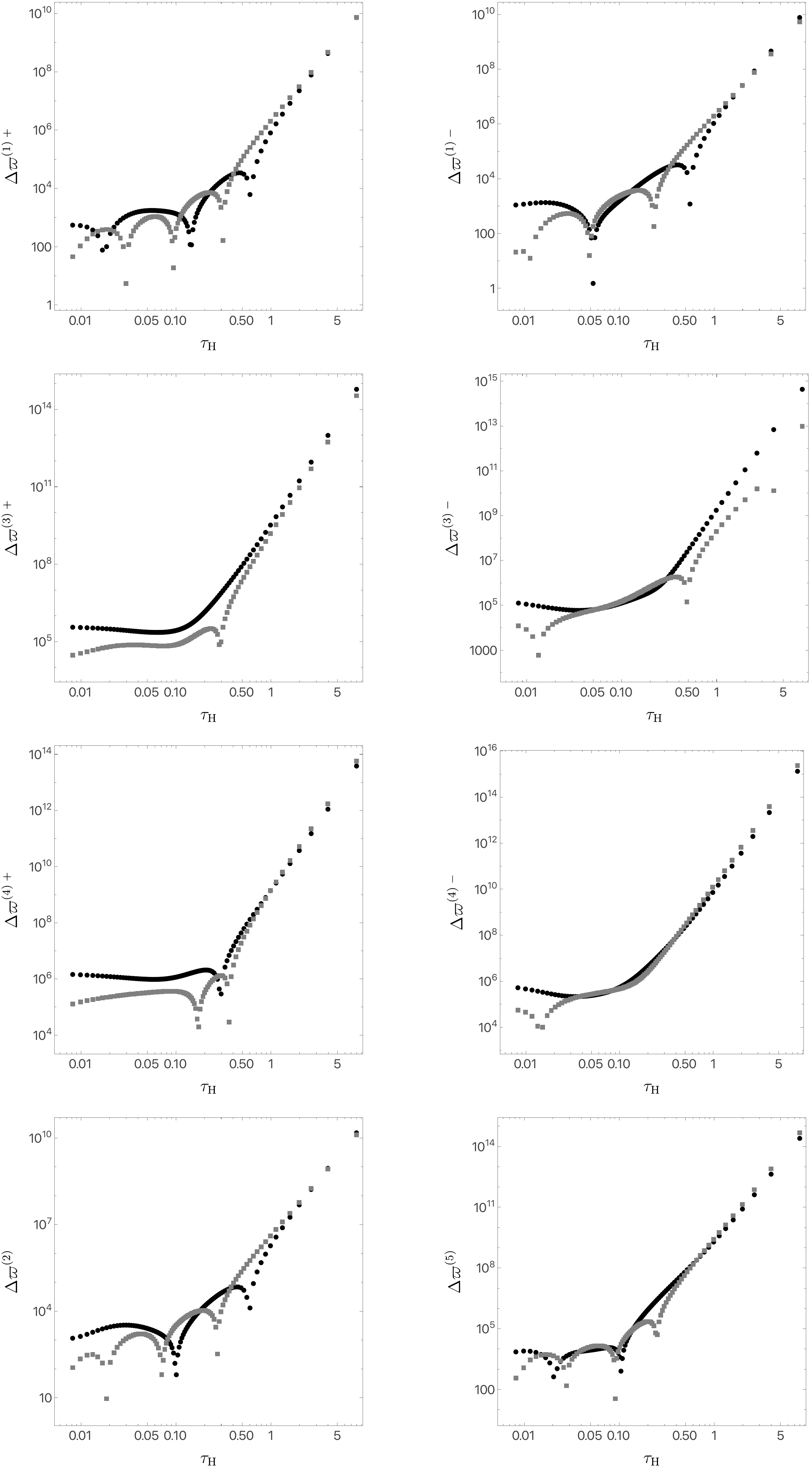}
    \caption{\label{fig:data1CA}
$\left|\mathrm{Re}\,\Delta\varpi^{(k)\pm}_{220}\right|$ (black disks) and 
$\left|\mathrm{Im}\,\Delta\varpi^{(k)\pm}_{220}\right|$ (grey squares) as functions of 
$\tau_H$ ($\log$-$\log$ scale). The different sectors are indicated on the right of each plot. The data are shown in the grand-canonical ensemble, where the EFT black holes and 
the Kerr black holes are compared at fixed temperature and angular velocity.}
\end{figure*}

In this section of the Supplemental Material we present the data for the remaining EFT coefficients. In general, the quartic deformations produce larger frequency shifts than their cubic counterparts (see Fig.~\ref{fig:data1A}). Moreover, while the cubic EFT deformations already display clear discrete self‑similar behaviour, for the quartic case only the odd‑parity sector exhibits a pronounced signal (see Fig.~\ref{fig:data1CA}).

\twocolumngrid

\end{document}